\documentclass[12pt,draftcls,onecolumn]{IEEEtran}
\usepackage{multirow}
\usepackage{amssymb}
\usepackage[dvips]{graphicx}
\usepackage{amsmath}
\usepackage{amsthm}
\usepackage{latexsym,bm}
\usepackage{color}
\usepackage{subfigure}
\usepackage{longtable}
\usepackage{accents}
\usepackage{cite}
\usepackage{enumerate}
\usepackage{arydshln}
\usepackage{slashbox}
\usepackage{algorithmic}
\usepackage{float}
\usepackage{stfloats}

\makeatletter
\renewcommand{\maketag@@@}[1]{\hbox{\m@th\normalsize\normalfont#1}}%
\makeatother

\def\bB{{\mathbf{B}}}

\makeatletter
\def\widebar{\accentset{{\cc@style\underline{\mskip10mu}}}}
\def\Widebar{\accentset{{\cc@style\underline{\mskip8mu}}}}
\makeatother

\theoremstyle{plain}

\theoremstyle{definition}
\theoremstyle{definition}

\usepackage{lineno}
\begin{document}

\title{Design of Protograph LDPC-Coded MIMO-VLC \\~~Systems with Generalized Spatial Modulation
\thanks{L. Dai, and Y. Fang are with the School of Information Engineering, Guangdong University of Technology, Guangzhou 510006,
China (email: bs0109dl@163.com; fangyi@gdut.edu.cn).}
\thanks{Y. L. Guan is with the School of Electrical and Electronic Engineering,
Nanyang Technological University, Singapore (e-mail: eylguan@ntu.edu.sg).}
\thanks{M.~Guizani is with the Machine Learning, Mohamed Bin Zayed University of Artificial Intelligence (MBZUAI), Abu Dhabi, UAE (e-mail: mguizani@ieee.org).}
}
\author{Lin Dai, Yi Fang, {\em Member, IEEE}, Yong Liang Guan, {\em Senior Member, IEEE}, Mohsen Guizani, {\em Fellow, IEEE}\vspace{-0.5cm}}

\maketitle
\begin{abstract}
This paper investigates the bit-interleaved coded generalized spatial modulation (BICGSM) with iterative decoding (BICGSM-ID) for multiple-input multiple-output (MIMO) visible light communications (VLC). In the BICGSM-ID scheme, the information bits conveyed by the signal-domain (SiD) symbols and the spatial-domain (SpD) light emitting diode (LED)-index patterns
are coded by a protograph low-density parity-check (P-LDPC) code. 
Specifically, 
we propose a signal-domain symbol expanding and re-allocating (SSER) method for constructing a type of novel generalized spatial modulation (GSM) constellations, referred to as {\em SSERGSM constellations}, so as to boost the performance of the BICGSM-ID MIMO-VLC systems.
Moreover, by applying a modified PEXIT (MPEXIT) algorithm, we further design a family of rate-compatible P-LDPC codes, referred to as {\em enhanced accumulate-repeat-accumulate (EARA) codes}, which possess both excellent decoding thresholds and linear-minimum-distance-growth property. Both analysis and simulation results illustrate that the proposed SSERGSM constellations and P-LDPC codes can remarkably improve the convergence and decoding performance of MIMO-VLC systems. Therefore, the proposed P-LDPC-coded SSERGSM-mapped BICGSM-ID configuration is envisioned as a promising transmission solution to satisfy the high-throughput requirement of MIMO-VLC applications.
\end{abstract}


\section{Introduction}\label{sect:section-1}
As a promising complementary solution to alleviate the shortage of wireless spectrum resources, optical wireless communication (OWC) is an attractive technology which has drawn a great deal of attention from both the academic and industrial communities \cite{8960379}. Particularly, visible light communication (VLC) technology has been standardized in IEEE $802.15.7$ \cite{6016195}, and hence could become a commercialized method for indoor communication scenarios due to the low infrastructure cost. VLC technology enjoys the characteristic that lighting and communication can coexist without causing interference for each other \cite{8240590}. Moreover, advantages in the terms of energy efficiency, wide bandwidth, license-free application and higher security make VLC a promising solution to be part of the 5G wireless networks \cite{9060998}. In indoor VLC systems, the light emitting diodes (LEDs) are the main medium used for data transmission by modulating the intensity of the emitted light because of their superior switching capabilities. However, the LEDs have quite low modulation bandwidth (i.e., about $20$~MHz) \cite{7842427}, which greatly limits the transmission rate of the VLC systems. Meanwhile, the conventional transmission framework typically adopts single-input-single-output (SISO) technique, which cannot provide spatial diversity gain. To deal with this problem, some technologies such as multiple-input multiple-output (MIMO) \cite{8047677,8233180} have been applied to the VLC systems.

The conventional MIMO technology activates all LED emitters at every time slot and transmits the same information simultaneously, which results in the inter-channel interference (ICI). For averting ICI, a multiple-LED modulation scheme, called {\em spatial modulation (SM)}, has been proposed in \cite{5722087,8642337}. However, SM would lead to low spectral efficiency (SE) because it does not efficiently utilize the transmitted resources. Therefore, some other schemes such as generalized spatial modulation (GSM) \cite{7416970}, quad-LED complex modulation (QCM) \cite{7564640} and dual-LED complex modulation (DCM) \cite{8990446} have been developed. As a spectral-efficient multiple-input multiple-output visible light communication (MIMO-VLC) transmission scheme, the conventional GSM (ConGSM) simultaneously activates multiple LEDs and every activated LED is allowed to carry the different unipolar pulse amplitude modulated (UPAM) intensity level during each transmission instant \cite{7416970}. However, due to the channel correlation of the static MIMO-VLC channels under the line-of-sight (LOS) condition, the signal-domain (SiD) symbol design for the ConGSM is not the optimal.
In \cite{7902182}, a new SiD symbol-design method for {\em activate-space collaborative constellation GSM (ASCCGSM)} has been proposed by minimizing the power of the SiD symbols to improve the error performance of MIMO-VLC system. 
Beyond that, it is well known that the coded modulation (CM) scheme \cite{771140, 5508985, 6362672, 9373632} plays an important role in improving the overall system performance.


In \cite{5508985}, a trellis coded spatial modulation (TCSM) scheme was proposed, in which only SiD bits are protected by a trellis code to improve the performance over correlated channels. After that, the authors in \cite{6213919} have extended the TCSM scheme to an enhanced coded SM (ECSM) scheme which jointly encodes the bits conveyed in the spatial-domain (SpD) and SiD so as to achieve a higher coding gain in the indoor OWC systems. Furthermore, it has been pointed out in \cite{7023130} that a trellis coded GSM (TGCSM) scheme is spectrally efficient and robust against the correlated MIMO channels. Additionally, a bit-interleaved coded modulation (BICM) combined with SM has been proposed as a promising solution, which provides performance improvements against the channel correlation effect \cite{6362672}.
However, the BICM is not the optimal transmission scheme due to the independent demapping and decoding procedures. To compensate for the performance loss arising from separated demapping and decoding, iterative decoding (ID) is introduced at the receiver to enhance the BICM performance \cite{8883091}.
In this paper, we consider the transmission scheme for BICM combined with GSM and ID, referred to as {\em BICGSM-ID}.
Nevertheless, there exist some challenges in the design of the bit-interleaved coded generalized spatial modulation with iterative decoding (BICGSM-ID) scheme for the MIMO-VLC systems.

In communication systems, error-correction codes (ECCs) play a critical role in improving the system performance.
In \cite{6213919}, the convolutional codes have been applied in the ECSM-aided indoor OWC systems. The non-binary low-density parity-check (LDPC) codes and polar codes have been presented in coded spatial modulation (CSM) scheme in the SM-MIMO systems \cite{8290977,9387135}. As a class of capacity-approaching codes, protograph low-density parity-check (P-LDPC) codes \cite {8269289} have been widely exploited in the wireless communication systems due to their noticeable performance and simple implementation. Especially, the accumulate-repeat-by-$4$-jagged-accumulate (AR$4$JA) code \cite{5174517}, which possesses simple protograph structures to realize linear encoding and decoding, can achieve very desirable error performance in both low and high signal-noise-ratio (SNR)
regions over additive white Gaussian noise (AWGN) channels. However, the conventional P-LDPC codes may not perform well in specific BICM-ID VLC systems.

To facilitate the design and analysis of the P-LDPC-coded BICM systems, a variety of protograph-based extrinsic information transfer (PEXIT) algorithms \cite{6133952,7907160,8740906,8708250} have been developed to predict the system performance in the low SNR region.
On the other hand, in \cite{5174517,7112076}, the asymptotic weight distribution (AWD) has been formulated to estimate the typical minimum distance ratio (TMDR), i.e., the linear-minimum-distance growth property, of a P-LDPC code so as to predict the system performance in the high SNR region.
Although the performance of the P-LDPC-coded BICM systems has been intensively studied, there is still a lack of design methodology for P-LDPC-coded BICGSM-ID scheme, especially in the MIMO-VLC systems.

Driven by the aforementioned motivations, we investigate the performance of the P-LDPC-coded BICGSM-ID MIMO-VLC systems. The main contributions are in two-fold and can be summarized as follows. 
$1$) A type of novel GSM constellations, referred to as {\em symbol expanding and re-allocating GSM (SSERGSM)
constellations}, is proposed through an SSER method. Meanwhile, we utilize the constellation-constrained average mutual
information (AMI) to evaluate the achievable capacity of the proposed SSERGSM constellations in the MIMO-VLC systems.
$2$) Based on the modified protograph extrinsic information transfer (MPEXIT) algorithm, we put forward a novel family of rate-compatible P-LDPC codes, called {\em enhanced accumulate-repeat-accumulate
(EARA) codes}. The proposed EARA codes can not only
benefit from desirable decoding thresholds, but also possess the linear-minimum-distance-growth property in the BICGSM-ID MIMO-VLC systems. Simulations are carried out to verify that the proposed P-LDPC-coded BICGSM-ID schemes with the SSERGSM constellations outperform the state-of-the-art counterparts in the MIMO-VLC systems.

The remainder of this paper is organized as follows. Section~II introduces the P-LDPC-coded BICGSM-ID MIMO-VLC system model and presents the information-theoretic methodology
to calculate the AMIs of the GSM constellations. This section also describes the proposed an MPEXIT algorithm for the PLDPC-coded BICGSM-ID MIMO-VLC system.
Section~III proposes the SSERGSM constellation design, in which a new SSER method is exploited to design the SiD symbols. Section~IV constructs a new family of rate-compatible P-LDPC codes. Section~V provides various simulated results and discussions. Finally, Section~VI concludes the contributions of this paper.

{\em Notations}: Boldface lowercase and uppercase letters denote vectors and matrices, respectively. Let $(\cdot)^{\rm T}$, $\lfloor\cdot\rfloor$, $\|\cdot\|$ and $!$ denote the transpose operation, floor function, Frobenius norm and factorial operation, respectively. Besides, $|\cdot|$ represents the cardinality of a set.

\begin{figure*}[t]
  \center
  \includegraphics[width=6.5in,height=1.45in]{{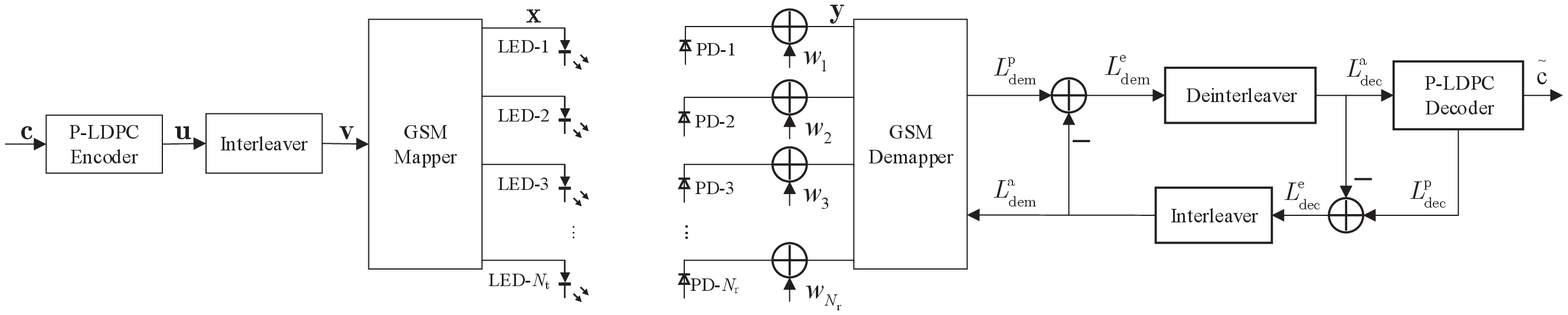}}
  \vspace{-2mm}
  \caption{Block diagram of a P-LDPC-coded BICGSM-ID MIMO-VLC system.}
  \label{fig:MIMO-VLC}
  \vspace{-5mm}
\end{figure*}

\section{System Model and Preliminaries}\label{sect:section-2}
\subsection{Conventional GSM}\label{sect:section-2-A}
In a conventional GSM (ConGSM) \cite{7416970}, we assume $N_{\rm t}$ and $N_{\rm a}$ (i.e., $N_{\rm a}\geq2$) are the numbers of transmit LEDs and active LEDs at each
transmission instant, respectively. It is apparent that the number of all possible LED activation patterns is $\Delta = N_{\rm t}!/(N_{\rm t}-N_{\rm a})!N_{\rm a}!$.
To ensure that the number of LED activation patterns is a power of two, $2^{\rho_d}$ effective LED activation patterns are selected for the $M$-ary unipolar PAM ($M$-UPAM) intensity level transmission, where $\rho_d= \lfloor\log_2\Delta\rfloor$.
The $M$-UPAM intensity levels can be formulated as \cite{7416970,8264686}
\begin{equation}
I_t = (2I_{\rm a}t)/(M+1), \forall t = 1,2,\cdots,M,
\label{eq:1}
\end{equation}where $I_{\rm a}$ is the average intensity level of each LED.

Consequently, there are two different types of constellations, i.e., spatial-domain (SpD) constellation and signal-domain (SiD) constellation, involved in GSM. The SpD constellation set $\Omega$ is dependent on the effective LED activation patterns (i.e., $|\Omega| = 2^{\rho_d}$) and the SiD constellation set $\Lambda$ is dependent on the order $M$ of UPAM and $N_{\rm a}$ (i.e., $|\Lambda| = M \cdot N_{\rm a}$).
As a result, at each transmission instant, the number of coded bits can be written as

\begin{equation}
\rho = \rho_d + {N_{\rm a}}\cdot\log_2M.
\label{eq:2}
\end{equation}
More details of the ConGSM can be referred to \cite{7416970}.

\subsection{System Model}\label{sect:section-2-B}
The block diagram of a P-LDPC-coded BICGSM-ID MIMO-VLC system is shown in Fig.~\ref{fig:MIMO-VLC}, in which an indoor LOS MIMO-VLC link with $N_{\rm t}$ LEDs and $N_{\rm r}$ photodetectors (PDs) at the transmitter and receiver, respectively, are considered. In such a scenario, we assume that $N_{\rm a}$ (i.e., $N_{\rm a}\geq2$) is the number of active LEDs at each
transmission instant.
In Fig.~\ref{fig:MIMO-VLC}, the input information-bit stream ${\bf{c}} = ({{\bf c}_1},{\bf c}_2,\ldots,{\bf c}_k )$ is first encoded to a length-$n$ codeword $\bf{u}$ (i.e., coded-bit stream) by exploiting a P-LDPC code. Further, the code-bit stream ${\bf{u}} = ({\bf u}_1,{\bf u}_2,\ldots,{\bf u}_n )$ is permuted by a random interleaver. Subsequently, every $\rho$ consecutive coded bits are processed by a GSM mapper. More specifically, the $\rho$ consecutive coded bits are divided into two parts in a sequential order. The first part including $\rho_{\rm d}$ bits is used to select the effective LED activation patterns, while the second part contains $\rho_{\rm s} = {N_{\rm a}}\log_2M $ bits, and every $\log_2M$ component bits are independently mapped to an $M$-UPAM intensity level.



After the GSM mapper, every $\rho$ (i.e., $\rho = \rho_{\rm d} + \rho_{\rm s}$) coded bits are carried by a GSM signal, and thus a length-$(n/\rho)$ GSM symbol sequence ${\bf \hat{v}} =
(\hat{\bf{v}}_1,\hat{\bf{v}}_2,\cdots,\hat{\bf{v}}_k,\cdots,\hat{\bf{v}}_{n/\rho})$ can be yieled.
Finally, each GSM symbol is converted into a transmission vector $\mathbf{x} = [x_1,x_2,\cdots,x_{N_{\rm t}}]^\mathbf{T}$, which is propagated through a MIMO-VLC channel.
The received signal can be given as
\begin{equation}
\mathbf{y} = \mathbf{Hx} + \mathbf{w},
\label{eq:3}
\end{equation}
where $\mathbf{y}=[y_1,y_2,\cdots,y_{N_{\rm r}}]^\mathbf{T}$ is a vector with size of $N_{\rm r}\times1$, and $\mathbf{x} = [x_1,x_2,\cdots,x_{N_{\rm t}}]^\mathbf{T}$ is the transmitted GSM signal vector. In particular, $\mathbf{x}$ consists of the $N_{\rm t}-N_{\rm a}$ zero elements
representing the inactive LEDs, and the $N_{\rm a}$ positive, real-valued elements choosing from the UPAM intensity symbol set.
Moreover, $\mathbf{w}$ is the noise vector with size of $N_{\rm r}\times1$, each of which can be modeled as the independent and identically distributed (i.i.d) additive white Gaussian noise (AWGN) with zero mean and variance ${\sigma^2 = N_0/2}$, where $N_0$ is the single sided noise power spectral density.
$\mathbf{H}$ is the LOS channel gain matrix with size of $N_{\rm r}\times N_{\rm t}$, and the entries of the $\mathbf{H}$ matrix can be determined by \cite{6213919}
\begin{equation}
h_{ij} = \left \{
\begin{array}{ll}
\frac{\varepsilon(\eta+1)A}{2 \pi d^2_{ij}} \cos^\eta(\phi_{ij})\cos(\psi_{ij}), & \; 0\leq\psi_{ij}\leq\Psi_{1/2}\\
0, & \; \psi_{ij}>\Psi_{1/2},
\end{array}
\right.
\label{eq:4}
\end{equation}
where $d_{ij}$ represents the distance between the $j$-th transmit LED and the $i$-th receive PD, $i\in(1,2,\cdots,N_{\rm r})$ and $j\in(1,2,\cdots,N_{\rm t})$. $\phi_{ij}$ and $\psi_{ij}$ are the angle of irradiance and incidence from the $j$-th transmit LED to the $i$-th receive PD, respectively. $\eta = - \ln2/\ln(\cos\Phi_{1/2})$ represents the Lambert's mode number, where $\Phi_{1/2}$ is the transmit semi-angle of the LED. Also, $\varepsilon$, $A$ and $\Psi_{1/2}$ represent the PD responsivity, PD area and half-power field-of-view (FOV) angle of the PD, respectively. In this paper, the average optical signal-to-noise ratio (OSNR) is defined as follows \cite{7500474,7915761}
\begin{equation}
{ \textmd{OSNR}} = P_{\rm rx}/(\sqrt{2R\rho}\sigma),
\label{eq:5}
\end{equation}where $R$ denotes the code rate and $P_{\rm rx} = \frac{1}{N_{\rm r}}\sum_{i=1}^{N_{\rm r}}\sum_{j=1}^{N_{\rm t}}{h_{ij}I}$ is the average received optical power, and $I$ is the average optical intensity being emitted.

At the receiver, each received GSM symbol is processed by a serially concatenated ID framework, which includes a GSM demapper and a P-LDPC decoder.
Specifically, given the $k$-th received symbol ${\mathbf y}_k$ and its corresponding bit sequence $\hat{\bf{v}}_k = (v_1^k,v_2^k,\cdots,v_{\rho_{\rm d}}^k,\cdots,v_\rho^k)$, in which the first consecutive $\rho_{\rm d}$ bits are used for selecting effective LED activation patterns with the set $\Omega$ and the remaining $\rho_{\rm s}$ bits are mapped to the SiD constellation with the set $\Lambda$,
we conveniently denote the SpD and SiD bit sequences as $\hat{\bf{v}}_{{\rm d},k} = (v_{{\rm d},1}^k,v_{{\rm d},2}^k,
\cdots,v_{{\rm d},l}^k,\cdots,v_{{\rm d},\rho_{\rm d}}^k)$ and
$\hat{\bf{v}}_{{\rm s},k} = (v_{{\rm s},1}^k,v_{{\rm s},2}^k,\cdots,v_{{\rm s},q}^k,
\cdots,v_{{\rm s},\rho_{\rm s}}^k)$, respectively.
Based on the $k$-th received symbol and its corresponding {\em a-priori} log-likelihood-ratios (LLRs) $L_{\rm dem}^{\rm a}$, the corresponding {\em a-posteriori} LLRs $L_{\rm dem}^{\rm p}$ output from the GSM demapper can be derived by \cite{123456}
\begin{equation}
\begin{aligned}
L_{\rm dem}^{\rm p}(v_{{\rm d},l}^k)=& L_{\rm dem}^{\rm a}(v_{{\rm d},l}^k) + \max_{{\bf x}\in\Lambda_q^b,{\bf x}\in\Omega_l^0}[-\|{\bf y}-{\bf Hx}\|^2/2\sigma^2 \\
& +Q_1] - \max_{{\bf x}\in\Lambda_q^b,{\bf x}\in\Omega_l^1}[-\|{\bf y}-{\bf Hx}\|^2/2\sigma^2 + Q_1],
\end{aligned}
\label{eq:6}
\end{equation}
\begin{equation}
\begin{aligned}
L_{\rm dem}^{\rm p}(v_{{\rm s},q}^k) =& L_{\rm dem}^{\rm a}(v_{{\rm s},q}^k) + \max_{{\bf x}\in\Omega_l^b,{\bf x}\in\Lambda_q^0}[-\|{\bf y}-{\bf Hx}\|^2/2\sigma^2  \\
& + Q_2] - \max_{{\bf x}\in\Omega_l^b,{\bf x}\in\Lambda_q^1}[-\|{\bf y}-{\bf Hx}\|^2/2\sigma^2 + Q_2],
\end{aligned}
\label{eq:7}
\end{equation}
where $l = 1,2,\cdots,\rho_{\rm d}$, $q = 1,2,\cdots,\rho_{\rm s}$, $Q_1 = \sum_{\tau = 1,\tau\neq l}^{\rho}(1-v_\tau^k)L_{\rm dem}^{\rm a}(v_\tau^k)$, and $Q_2 = \sum_{\tau = 1,\tau\neq {\rho_{\rm d}+q}}^{\rho}(1-v_\tau^k)L_{\rm dem}^{\rm a}(v_\tau^k)$. Additionally, $\Omega_l^b$ is the subset of $\Omega$ with the $l$-th labeling bit having value $b\in\{0,1\}$ and $\Lambda_q^b$ is the subset of $\Lambda$ with the $q$-th labeling bit having value $b\in\{0,1\}$.
Hence, the {\em extrinsic} LLRs $L_{\rm dem}^{\rm e}$ \big(i.e., $L_{\rm dem}^{\rm e} = L_{\rm dem}^{\rm p} - L_{\rm dem}^{\rm a}$\big) gleaned from the GSM demapper are deinterleaved and are viewed as the {\em a-priori} LLRs $L_{\rm dec}^{\rm a}$ into the P-LDPC decoder. Through a decoding procedure, the {\em extrinsic} LLRs $L_{\rm dec}^{\rm e}$ \big(i.e., $L_{\rm dec}^{\rm e} = L_{\rm dec}^{\rm p} - L_{\rm dec}^{\rm a}$\big) output from the P-LDPC decoder are re-interleaved and fed back as the updated {\em a-priori} LLRs $L_{\rm dem}^{\rm a}$ for the next iteration. In such an iterative demapping and decoding framework, the {\em extrinsic} LLRs and {\em a-priori} LLRs are iteratively updated between the demapper and the decoder so as to improve the reliability of the {\em a-posteriori} LLRs and enhance the MIMO-VLC system performance.

\subsection{Achievable Average Mutual Information}\label{sect:section-2-C}
Here, we discuss the AMI for the BICGSM transmission scheme in the MIMO-VLC systems. For the BICGSM scheme, the transmitted GSM symbol vector ${\bf x}$ is determined jointly by the effective LED activation pattern $d \in \Omega$  and the SiD constellation symbol $s\in\Lambda$, in which each symbol $s$ contains $N_{\rm a}$ UPAM intensity levels.

From the perspective of equivalent parallel channels, at each transmitted instant, the joint mapping for the $\rho$ bits in the set $\Omega$ and $\Lambda$
can be viewed as $\rho$ parallel binary
sub-channels which are independent and memoryless.
Suppose that the receiver has ideal channel state information (CSI). The AMI can reflect the maximum information rate for error-free transmission, and thus the AMI can be used as a criterion for evaluating the channel transmission performance.
We assume that $d$ and $s$ are the independent uniformly distributed random variables in the sets $\Omega$ and $\Lambda$, respectively. The spatial-domain AMI (SiD-AMI) and signal-domain AMI (SpD-AMI), denoted by $I_{{\rm SpD}}$ and $I_{{\rm SiD}}$ respectively, can be calculated as \cite{8237200}
\begin{equation}
\begin{aligned}
I_{{\rm SpD}} &= \sum_{i=1}^{\rho_{\rm d}} I(v_{{\rm d},i};{\bf y}|{\bf H})
\\
& =\rho_{\rm d} - \sum_{i=1}^{\rho_{\rm d}}\mathbb{E}_{{v_{{\rm d},i}},{\bf y},{\bf H}}
\left[\log_2\frac{\sum_{d\in\Lambda, s\in\Omega}p({\bf y}| d,s,{\bf H})}{\sum_{d\in\Lambda,s\in\Omega_i^{v_{{\rm d},i}}}p({\bf y}|d,s,{\bf H})}\right],
\end{aligned}
\end{equation}
\begin{equation}
\begin{aligned}
I_{{\rm SiD}} &= \sum_{j=1}^{\rho_{\rm s}} I(v_{{\rm s},j};{\bf y}|{\bf H}) \\
&=\rho_{\rm s} -\sum_{j = 1}^{\rho_{\rm s}}\mathbb{E}_{{v_{s,j}},{\bf y},{\bf H}}
\left[\log_2\frac{\sum_{d\in\Lambda, s\in\Omega}p({\bf y}|d,s,{\bf H})}{\sum_{ s\in\Omega, d\in\Lambda_i^{v_{{\rm s},j}}}p({\bf y}|d,s,{\bf H})}\right],
\end{aligned}
\end{equation}
where $v_{{\rm d},i}$ $(i = 1, 2,\cdots, \rho_{\rm d})$ and $v_{{\rm s},j}$ $(j = 1, 2,\cdots, \rho_{\rm s})$ are the $i$-th SpD
labeling bit and the $j$-th SiD labeling bit, respectively. Moreover, $\Omega_i^{v_{{\rm d},i}}$ and $\Lambda_j^{v_{{\rm s},j}}$ are the subset of $\Omega$ with the $i$-th bit $v_{{\rm d},i}$ $\left(v_{{\rm d},i}\in\{{0,1}\}\right)$ and the subset of $\Lambda$ with the $j$-th bit $v_{{\rm s},j}$ $\left(v_{{\rm s},j}\in\{{0,1}\}\right)$, respectively. Additionally, $p({\bf y}|d,s,{\bf H})$ denotes the probability density function (PDF) of the receiver vector ${\bf y}$ conditioned on the effective LED activation pattern $d \in \Omega$ and the SiD
constellation symbol $s\in\Lambda$.
Therefore, the constrained AMI for a BICGSM scheme in the MIMO-VLC systems, called {\em BICGSM-AMI} is given by
\begin{equation}
I_{{\rm BICGSM}} = I_{{\rm SpD}} + I_{{\rm SiD}}.
\end{equation}

\subsection{Modified PEXIT Algorithm}
To the best of our knowledge, PEXIT algorithm can be used to track the evolution of the mutual-information between the variable-node (VN) decoder and the check-node (CN) decoder \cite{8269289}. Through such a manner, PEXIT algorithm is able to evaluate the decoding threshold of a P-LDPC code so as to reveal the error performance at low SNR region under an iterative decoder. By modifying the traditional PEXIT algorithm \cite{4411526}, one can evaluate the convergence performance of the P-LDPC-coded BICGSM-ID scheme in MIMO-VLC systems.


\begin{table}[tbp]\scriptsize
\center
\caption{Mapping table of the ConGSM constellation with $N_{\rm t} = N_{\rm r} = 4$, $N_{\rm a} = 2$ and $\rho = 4$ (i.e., $M = 2$).}
\begin{tabular}{|c|c|c|c|}
\hline
\multirow{2}{*}{Label}   &  Effective LED     &  SiD modulated   & Transmitted   \\
         &activation pattern &symbol &vector $\mathbf{x}$\\
\hline
$0000$   &  $(1,2)$    &  $(2I_{\rm a}/3,2I_{\rm a}/3)$ & $[2I_{\rm_a}/3,2I_{\rm a}/3,0,0]^{\rm T}$\\
\hline
$0001$   &  $(1,2)$    &  $(2I_{\rm a}/3,4I_{\rm a}/3)$ & $[2I_{\rm a}/3,4I_{\rm a}/3,0,0]^{\rm T}$\\
\hline
$0011$   &  $(1,2)$    &  $(4I_{\rm a}/3,4I_{\rm a}/3)$ & $[4I_{\rm a}/3,4I_{\rm a}/3,0,0]^{\rm T}$\\
\hline
$0010$   &  $(1,2)$    &  $(4I_{\rm a}/3,2I_{\rm a}/3)$ & $[4I_{\rm a}/3,2I_{\rm a}/3,0,0]^{\rm T}$\\
\hline
$0100$   &  $(1,3)$    &  $(2I_{\rm a}/3,2I_{\rm a}/3)$ & $[2I_{\rm a}/3,0,2I_{\rm a}/3,0]^{\rm T}$\\
\hline
$0101$   &  $(1,3)$    &  $(2I_{\rm a}/3,4I_{\rm a}/3)$ & $[2I_{\rm a}/3,0,4I_{\rm a}/3,0]^{\rm T}$\\
\hline
$0111$   &  $(1,3)$    &  $(4I_{\rm a}/3,4I_{\rm a}/3)$ & $[4I_{\rm a}/3,0,4I_{\rm a}/3,0]^{\rm T}$\\
\hline
$0110$   &  $(1,3)$    &  $(4I_{\rm a}/3,2I_{\rm a}/3)$ & $[4I_{\rm a}/3,0,2I_{\rm a}/3,0]^{\rm T}$\\
\hline
$1000$   &  $(1,4)$    &  $(2I_{\rm a}/3,2I_{\rm a}/3)$ & $[2I_{\rm a}/3,0,0,2I_{\rm a}/3]^{\rm T}$\\
\hline
$1001$   &  $(1,4)$    &  $(2I_{\rm a}/3,4I_{\rm a}/3)$ & $[2I_{\rm a}/3,0,0,4I_{\rm a}/3]^{\rm T}$\\
\hline
$1011$   &  $(1,4)$    &  $(4I_{\rm a}/3,4I_{\rm a}/3)$ & $[4I_{\rm a}/3,0,0,4I_{\rm a}/3]^{\rm T}$\\
\hline
$1010$   &  $(1,4)$    &  $(4I_{\rm a}/3,2I_{\rm a}/3)$ & $[4I_{\rm a}/3,0,0,2I_{\rm a}/3]^{\rm T}$\\
\hline
$1100$   &  $(2,3)$    &  $(2I_{\rm a}/3,2I_{\rm a}/3)$ & $[0,2I_{\rm a}/3,2I_{\rm a}/3,0]^{\rm T}$\\
\hline
$1101$   &  $(2,3)$    &  $(2I_{\rm a}/3,4I_{\rm a}/3)$ & $[0,2I_{\rm a}/3,4I_{\rm a}/3,0]^{\rm T}$\\
\hline
$1111$   &  $(2,3)$    &  $(4I_{\rm a}/3,4I_{\rm a}/3)$ & $[0,4I_{\rm a}/3,4I_{\rm a}/3,0]^{\rm T}$\\
\hline
$1110$   &  $(2,3)$    &  $(4I_{\rm a}/3,2I_{\rm a}/3)$ & $[0,4I_{\rm a}/3,2I_{\rm a}/3,0]^{\rm T}$\\
\hline
\end{tabular}\vspace{-2mm}
\label{tab:tab1}
\end{table}


To facilitate the treatment of the modified PEXIT (MPEXIT) algorithm for the P-LDPC-coded BICGSM-ID scheme, we first introduce the concept of P-LDPC codes.
A protograph can be seen as a Tanner graph \cite{Thorpe2003Low} containing a few variable nodes, check nodes and edges. In the Tanner graph, each edge connects a variable
node and a check node and the parallel edges are allowed. Assume that a protograph possesses $n_v$ variable nodes and $n_c$ check nodes, which can be represented by an $n_c\times n_v$ base matrix $\mathbf{B} = (b_{i,j})$, $b_{i,j}$ is the number of edges connecting $C_i$ with $V_j$. Furthermore, a larger protograph graph corresponding to a P-LDPC code can be obtained from a given protograph (resp. base matrix) by a ``copy-and-permute (lifting)'' operation which is usually implemented by a modified progressive-edge-growth (PEG) algorithm \cite{555555}.

In the BICGSM-ID  MIMO-VLC system, Suppose that $G_{1}$ is the maximum number of inner iteration for the P-LDPC decoder while $G_{2}$ is the maximum number of outer iteration between the GSM demapper and the P-LDPC decoder. The MPEXIT algorithm is shown as follows.
\begin{enumerate}
\item \textbf{Initialization:}
Given an OSNR, initialize the incoming {\em a-priori} MI from the $i$-th variable node to the $j$-th check node $I_{\rm av}(i,j) = 0$.
\item \textbf{Extrinsic MI calculation of GSM demapper:}
For the $k$-th channel output symbol $\hat{\bf{v}}_k$, and the corresponding the SpD bits $\hat{\bf{v}}_{{\rm d},k} = (v_{{\rm d},1}^k,v_{{\rm d},2}^k,
\cdots,v_{{\rm d},i}^k,\cdots,v_{{\rm d},\rho_{\rm d}}^k)$
and the SiD bits $\hat{\bf{v}}_{{\rm s},k}= (v_{{\rm s},1}^k,v_{{\rm s},2}^k,
\cdots,v_{{\rm s},i}^k,\cdots,v_{{\rm s},\rho_{\rm s}}^k)$,
the extrinsic MI $I_{\rm {e,dem}}^{\alpha,k}$ between $\hat{\bf{v}}_{{\rm \alpha},k}$ and $L_{\rm dem}^p(v_{\alpha}^k)$ in the GSM demapper can be calculated by Monte-Carlo simulation, where $L_{\rm dem}^{\rm p}(v_{\alpha}^k)$ is expressed by equations \eqref{eq:6} and \eqref{eq:7}, and $\alpha = \{{\rm d},{\rm s}\}$.

\item \textbf{Channel MI calculation of variable nodes:}
The channel MI of the $i$-th variable node $I_{\rm ch}(i)$ is calculated by
\begin{equation}
I_{\rm ch}(i) = \frac{1}{n}\sum_{k=1}^{n/\rho}(I_{\rm {e,dem}}^{{\rm d},k}\cdot\rho_{\rm d} + I_{\rm {e,dem}}^{{\rm s},k}\cdot\rho_{\rm s}).
\end{equation}
Specially, $I_{\rm ch} (i) = 0$ if the $i$-th variable node is punctured in a protograph.
\item \textbf{Extrinsic MI update between variable nodes and check nodes:}
In the inner iteration process, the {\em extrinsic} MIs are iteratively updated between the variable nodes and the check nodes, where the detailed process is given in \cite{4411526}.
\item \textbf{A-priori MI update for GSM demapper:}
Exploiting the {\em a-priori} MI passed from check nodes to variable nodes, {\em a-priori} MI $I_{\rm dem}^{\rm a}$ for the GSM dmapper can be measured by
\begin{equation}
I_{\rm dem}^{\rm a}(j) = J\bigg(\bigg(\sum_{i = 1}^{n_c}b_{i,j}[J^{-1}(I_{\rm vn}^{\rm a}(i,j))]^2\bigg)^{1/2}\bigg),
\label{eq:dem-a}
\end{equation}
where $I_{\rm vn}^{\rm a}(i,j)$ is the {\em a-priori} MI passed from the $j$-th check node to the $i$-th variable node, and $J(.)$ function is derived in \cite{8740906}.
Then $I_{\rm dem}^{\rm a}$ can be further used for the next outer iteration.
\end{enumerate}

Through this way, we can get the lowest OSNR value for which all {\em a-posteriori} MIs of the variable nodes converges to $1$ by repeating Steps $2)\sim 5)$ with different OSNR values.


{\em Remark $1$:} To ensure the accuracy of the evaluated result, the proposed MPEXIT algorithm assumes that a finite binary codeword including both $0$ and $1$ bits is transmitted when calculating the MI, while the traditional PEXIT algorithm assumes an infinite-length all-zero codeword to do so. In addition, the proposed MPEXIT algorithm considering the silent features of GSM and indoor MIMO-VLC channel has a different channel MI expression $I_{\rm ch}(i)$ from that in the traditional PEXIT algorithm.

\section{Design of Proposed SSERGSM for BICGSM-ID MIMO-VLC systems}\label{sect:section-3}
\begin{table}[tbp]\scriptsize
\center
\caption{Mapping table of the proposed SSERGSM constellation with $N_{\rm t} = N_{\rm r} = 4$, $N_{\rm a} = 2$ and $\rho = 4$ (i.e., $M = 2$).}
\begin{tabular}{|c|c|c|c|}
\hline
\multirow{2}{*}{Label}   &  Effective LED     &  SiD modulated   & Transmitted  \\

         & activation pattern & symbol & vector $\mathbf{x}$ \\
\hline
$0000$   &  $(1,2)$    &  $(I_{\rm a}/6,I_{\rm a}/6)$ & $[I_{\rm_a}/6,I_{\rm a}/6,0,0]^{\rm T}$\\
\hline
$0001$   &  $(1,2)$    &  $(I_{\rm a}/6,7I_{\rm a}/6)$ & $[I_{\rm a}/6,7I_{\rm a}/6,0,0]^{\rm T}$\\
\hline
$0011$   &  $(1,2)$    &  $(7I_{\rm a}/6,7I_{\rm a}/6)$ & $[7I_{\rm a}/6,7I_{\rm a}/6,0,0]^{\rm T}$\\
\hline
$0010$   &  $(1,2)$    &  $(7I_{\rm a}/6,I_{\rm a}/6)$ & $[7I_{\rm a}/6,I_{\rm a}/6,0,0]^{\rm T}$\\
\hline
$0100$   &  $(1,3)$    &  $(2I_{\rm a}/6,2I_{\rm a}/6)$ & $[2I_{\rm a}/6,0,2I_{\rm a}/6,0]^{\rm T}$\\
\hline
$0101$   &  $(1,3)$    &  $(2I_{\rm a}/6,8I_{\rm a}/6)$ & $[2I_{\rm a}/6,0,8I_{\rm a}/6,0]^{\rm T}$\\
\hline
$0111$   &  $(1,3)$    &  $(8I_{\rm a}/6,8I_{\rm a}/6)$ & $[8I_{\rm a}/6,0,8I_{\rm a}/6,0]^{\rm T}$\\
\hline
$0110$   &  $(1,3)$    &  $(8I_{\rm a}/6,2I_{\rm a}/6)$ & $[8I_{\rm a}/6,0,2I_{\rm a}/6,0]^{\rm T}$\\
\hline
$1000$   &  $(1,4)$    &  $(4I_{\rm a}/6,4I_{\rm a}/6)$ & $[4I_{\rm a}/6,0,0,4I_{\rm a}/6]^{\rm T}$\\
\hline
$1001$   &  $(1,4)$    &  $(4I_{\rm a}/6,10I_{\rm a}/6)$ & $[4I_{\rm a}/6,0,0,10I_{\rm a}/6]^{\rm T}$\\
\hline
$1011$   &  $(1,4)$    &  $(10I_{\rm a}/6,10I_{\rm a}/6)$ & $[10I_{\rm a}/6,0,0,10I_{\rm a}/6]^{\rm T}$\\
\hline
$1010$   &  $(1,4)$    &  $(10I_{\rm a}/6,4I_{\rm a}/6)$ & $[10I_{\rm a}/6,0,0,4I_{\rm a}/6]^{\rm T}$\\
\hline
$1100$   &  $(2,3)$    &  $(5I_{\rm a}/6,5I_{\rm a}/6)$ & $[0,5I_{\rm a}/6,5I_{\rm a}/6,0]^{\rm T}$\\
\hline
$1101$   &  $(2,3)$    &  $(5I_{\rm a}/6,11I_{\rm a}/6)$ & $[0,5I_{\rm a}/6,11I_{\rm a}/6,0]^{\rm T}$\\
\hline
$1111$   &  $(2,3)$    &  $(11I_{\rm a}/6,11I_{\rm a}/6)$ & $[0,11I_{\rm a}/6,11I_{\rm a}/6,0]^{\rm T}$\\
\hline
$1110$   &  $(2,3)$    &  $(11I_{\rm a}/6,5I_{\rm a}/6)$ & $[0,11I_{\rm a}/6,5I_{\rm a}/6,0]^{\rm T}$\\
\hline
\end{tabular}
\label{tab:tab2}
\end{table}

\begin{figure}[tbp]
\center
\includegraphics[width=2.8in,height=1.85in]{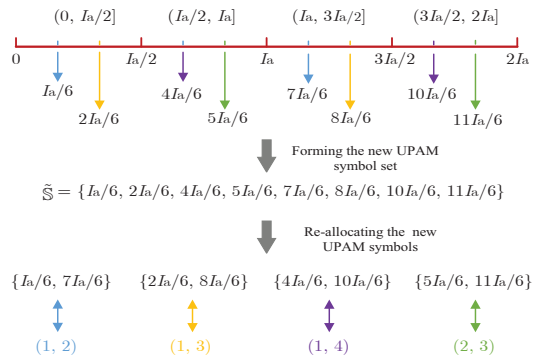}
\vspace{-3mm}
\caption{Design principle of the proposed SSERGSM constellation with $N_{\rm t} = N_{\rm r} = 4$, $N_{\rm a} = 2$ and $\rho = 4$ (i.e., $M = 2$).
\label{fig:SSER-principle}}
\vspace{-2mm} 
\end{figure}

In the MIMO-VLC systems, SiD modulated symbols are the critical parameter for formulating the GSM constellation, which can be designed and optimized so as to boost the system performance.


In the ConGSM, the SiD bits are modulated using the $M$-UPAM intensity levels for all effective LED
activation patterns. In other words, the intensity levels are selected from an $M$-UPAM symbol set $\mathbb{S} =\{I_1,I_2,\cdots,I_M\} $, the SpD bits allocated for all effective LED activation pattern share the same $M$-UPAM symbol set $\mathbb{S}$. As a example, the mapping table of ConGSM constellation for a $4\times4$ MIMO-VLC system with $N_{\rm a} = 2$ and $\rho = 4$ (i.e., $M = 2$) is shown in Table~\ref{tab:tab1}, where we choose $(1,2), (1,3), (1,4)$ and $(2,3)$ as the effective LED activation patterns, and the numbers $1$, $2$, $3$ and $4$ respectively represent the $1$-th, $2$-nd, $3$-rd and $4$-th LED indices.
Given a MIMO-VLC channel (i.e., a channel matrix $\mathbf{H}$), the correlation property of each effective LED activation pattern can be measured by the corresponding normalized instantaneous channel correlation matrix (NICCM) \cite{9076847}. The larger the value of elements in NICCM is, the stronger the correlation of the corresponding LED activation pattern possesses.
Due to the correlation property of the effective LED activation patterns, the ConGSM with all effective LED activation patterns sharing the same $M$-UPAM symbol set $\mathbb{S}$ may make the extrinsic MIs output from the demapper unreliable, which degrades the system performance.
Therefore, we put forward a signal-domain symbol expanding and re-allocating (SSER) method for constructing a type of novel GSM constellations to improve the performance of the BICGSM-ID MIMO-VLC system. The detailed SSER design method is described as follows.
\begin{enumerate}[1)]
\item
\textbf{Expanding the number of the UPAM symbols:}
 Given an $M$-ary UPAM symbol set ${\mathbb S} =\{I_1,I_2,\cdots,I_M\}$, which must satisfy $I_t\in(0,2I_{\rm a}], \forall t\in1,2,\cdots,M$, and the number of effective LED activation patterns $\beta = 2^{\rho_{\rm d}}$. First, we evenly divide the entire symbol space $(0,2I_{\rm a}]$ into $\beta$ symbol subspaces, i.e., $(2I_{\rm a}\tau/\beta,2I_{\rm a}(\tau+1)/\beta]$, $\forall \tau\in0,1,\cdots,\beta-1$. Subsequently, we choose $M$ UPAM symbols from each symbol subspace, and define the selection principle as
\begin{equation}
\tilde{I}_{n'+ M\tau} = \frac{2I_{\rm a}n'}{\beta(M+1)} + \frac{2I_{\rm a}\tau}{\beta},\;\; \forall \tau\in0,1,\cdots,\beta-1,
\end{equation}
where $n' = 1,2,\cdots,M$. Therefore, the new UPAM symbol set $\tilde{\mathbb{S}} =\{\tilde{I}_1,\tilde{I}_2,\cdots,\tilde{I}_{\beta M}\}$ is generated.
Taking the case of $\rho = 4$ (i.e., $M = 2$) as a example, and the detailed design principle are shown in Fig.~\ref{fig:SSER-principle}. Referring to this figure, one can obtain $4$ symbol subspaces, i.e., $(0,I_{\rm a}/2]$, $(I_{\rm a}/2,I_{\rm a}]$, $(I_{\rm a},3I_{\rm a}/2]$ and $(3I_{\rm a}/2,2I_{\rm a}]$. Then, we choose $2$ UPAM symbols from each symbol subspace
to form the new UPAM symbol set $\tilde{\mathbb{S}} = \{I_{\rm a}/6, 2I_{\rm a}/6, 4I_{\rm a}/6, 5I_{\rm a}/6, 7I_{\rm a}/6, 8I_{\rm a}/6, 10I_{\rm a}/6, 11I_{\rm a}/6\}$. Through this way, it is possible for all effective LED activation patterns to possess different UPAM symbol subsets.

\item
\textbf{Re-allocating the UPAM symbols for $\beta$ effective LED activation patterns:}
For the expanded $\beta M$ UPAM modulation symbols obtained from the first-step design, we first rank the $\beta M$ symbols in an ascending order of their corresponding intensity values. Then, we select $M$ different UPAM modulation symbols at maximum equal-intensity interval for each effective LED activation patterns. For example, the effective LED activation patterns $(1,2)$, $(1,3)$, $(1,4)$ and $(2,3)$ possess the UPAM symbol subsets $\{I_{\rm a}/6, 7I_{\rm a}/6\}$, $\{2I_{\rm a}/6, 8I_{\rm a}/6\}$, $\{4I_{\rm a}/6, 10I_{\rm a}/6\}$ and $\{5I_{\rm a}/6, 11I_{\rm a}/6\}$, respectively. Finally, the SiD bits in each effective LED activation pattern are mapped to the corresponding UPAM modulation symbols.

\end{enumerate}

Based on the above two steps, one can construct the SSERGSM constellation according to a given modulation order $M$ and the number of effective LED activation patterns $\beta$, which can significantly enhance the demapper reliability so as to ensure excellent BICGSM-ID performance. As an example, the mapping table of SSERGSM constellation for a $4\times4$ MIMO-VLC system with $N_{\rm a} = 2$ and $\rho = 4$ (i.e., $M = 2$) is shown in Table~\ref{tab:tab2}. To verify the superiority of the proposed SSERGSM constellation, the convergence performance of BICGSM-ID will be investigated in the forthcoming subsections.

\vspace{-1mm}
\subsection{Simulation Settings and BICGSM-AMI Analysis}\label{sect:section-3-B}
\subsubsection{Simulation Settings}\label{sect:section-3-B-1}
\begin{figure}[tbp]
\centering
\includegraphics[width=3.2in,height=1.75in]{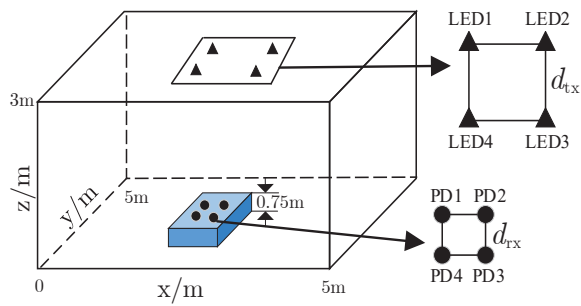}\vspace{-3mm}
\caption{Geometric setup of the channel model for a $4\times4$ indoor LOS MIMO-VLC system.}
\label{fig:simulation model}  \vspace{-5mm} 
\end{figure}

Unless otherwise mentioned, we consider a $4\times4$ indoor LOS MIMO-VLC channel and assume that the geometric setup of the channel model is shown in Fig~\ref{fig:simulation model}. In this figure, the LEDs and PDs are aligned in a quadratically $2\times2$ array which is located in the middle of a $5.0{\rm m}\times5.0{\rm m}\times3.0{\rm m}$ room, thus the four LEDs and four PDs construct two squares of lengths $d_{\rm tx}$ and $d_{\rm rx}$, respectively.
Moreover, the PDs are placed at the plane $z = 0.75{\rm m}$ (e.g., the height of a table). Besides, the parameters utilized for transceiver are summarized in Table~\ref{tab:tab3}.
Based on this model, we investigate different MIMO-VLC channel setups by varying the spacing between two neighboring LEDs (i.e., $d_{\rm tx}$), while the other system parameters remain unchanged. In this paper, the setting about the values of $d_{\rm tx}$ is the same as those in \cite{6213919}, i.e., $d_{\rm tx} = 0.3{\rm m},~0.5{\rm m},~{\rm and}~0.7{\rm m}$.
Note that three different values of $d_{\rm tx}$ represent three different channel matrices, and the smaller the value of $d_{\rm tx}$ is, the stronger the correlation of the corresponding channel is. 

\begin{table}[t]\scriptsize
\center
\caption{Parameter setting for the transceiver of an indoor MIMO-VLC system.}
\begin{tabular}{|c||c|c|}
\hline
\multirow{3}*{Room} &  {\rm Length, i.e.,} $x$ & $5{\rm m}$ \\
\cline{2-3} & {\rm Width, i.e.,} $y$	& $5{\rm m}$	     \\
\cline{2-3} & {\rm Height, i.e.,} $z$	& $3{\rm m}$	     \\
\hline
\multirow{5}*{Transmitter} & {\rm Number of LEDs, i.e.,} $N_{\rm t}$ & 4 \\
\cline{2-3}  & {\rm Number of the activated LEDs, i.e.,}~$N_{\rm a}$ & 2  \\
\cline{2-3}  & {\rm Height from the floor} & $3{\rm m}$ \\
\cline{2-3} & {\rm LED semi-angle, i.e.,} $\Phi_{1/2}$ & $8^{\rm o}$ \\
\cline{2-3} & $d_{\rm tx}$ & {$0.3{\rm m}, 0.5{\rm m}, 0.7{\rm m}$}  \\
\hline
\multirow{6}*{Receiver} & {\rm Number of PDs, i.e.,} $N_{\rm r}$ & 4 \\
\cline{2-3}  &  {\rm Height from the floor} & $0.75{\rm m}$ \\
\cline{2-3}  &  {\rm PD responsivity, i.e.,} $\varepsilon$ & $0.434{\rm A/W}$ \\
\cline{2-3}  &  {\rm PD area, i.e.,} $A$ & $7{\rm mm}^2$  \\
\cline{2-3}  &  {\rm FOV angle of PD, i.e.,} $\Psi_{1/2}$  & $55^{\rm o}$ \\
\cline{2-3}  &  $d_{\rm rx}$ & $0.1{\rm m}$ \\
\hline
\end{tabular}
\label{tab:tab3}
\end{table}

\subsubsection{BICGSM-AMI Analysis}\label{sect:section-3-B-2}
\begin{figure}[tbp]
\centering\vspace{-3mm}
\subfigure[\hspace{-0.5cm}]{\label{fig:dtx5-AMIa}
\includegraphics[width=2.8in,height=2.05in]{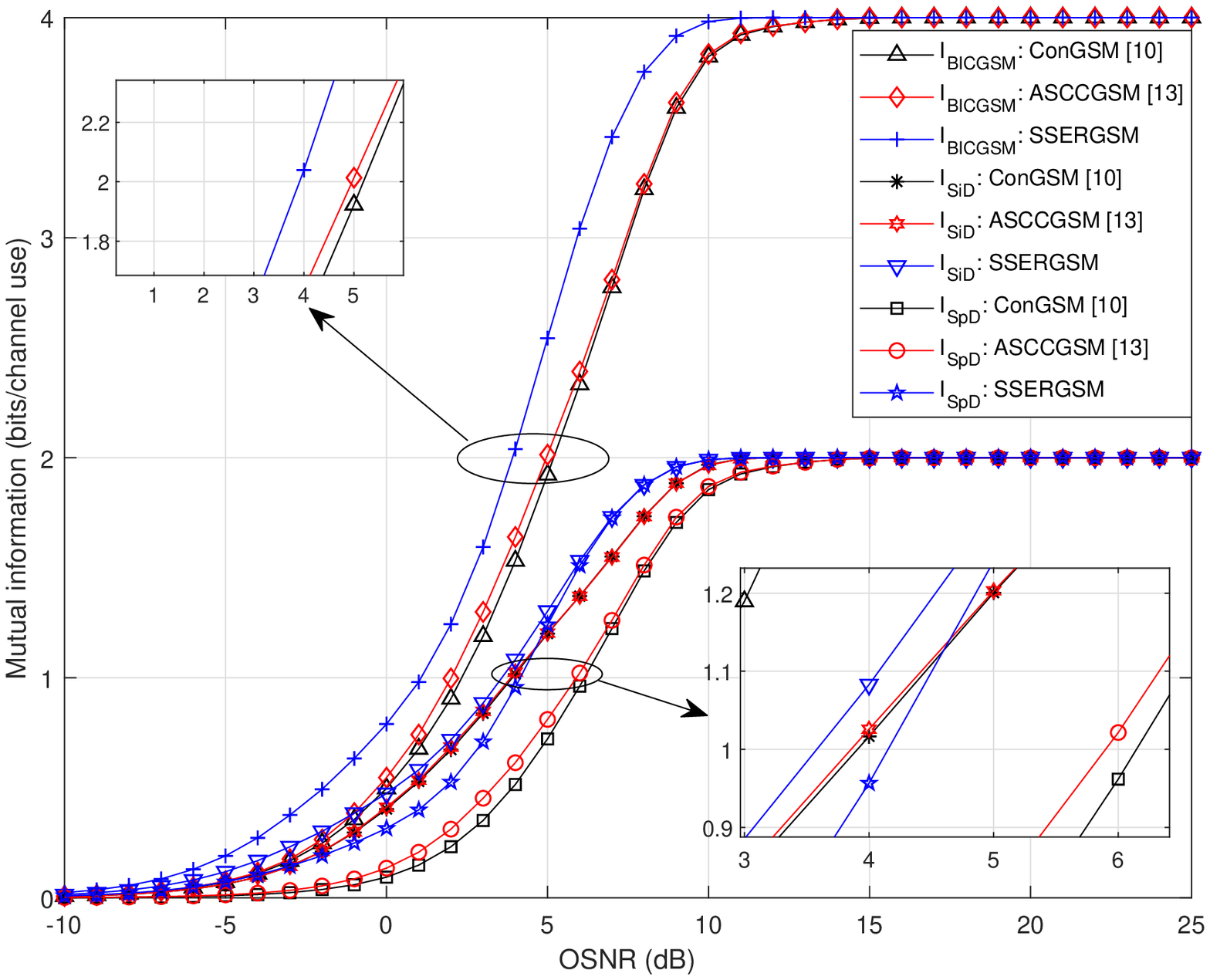}}\vspace{-2mm}
\subfigure[\hspace{-0.5cm}]{\label{fig:dtx5-AMIb}
\includegraphics[width=2.8in,height=2.05in]{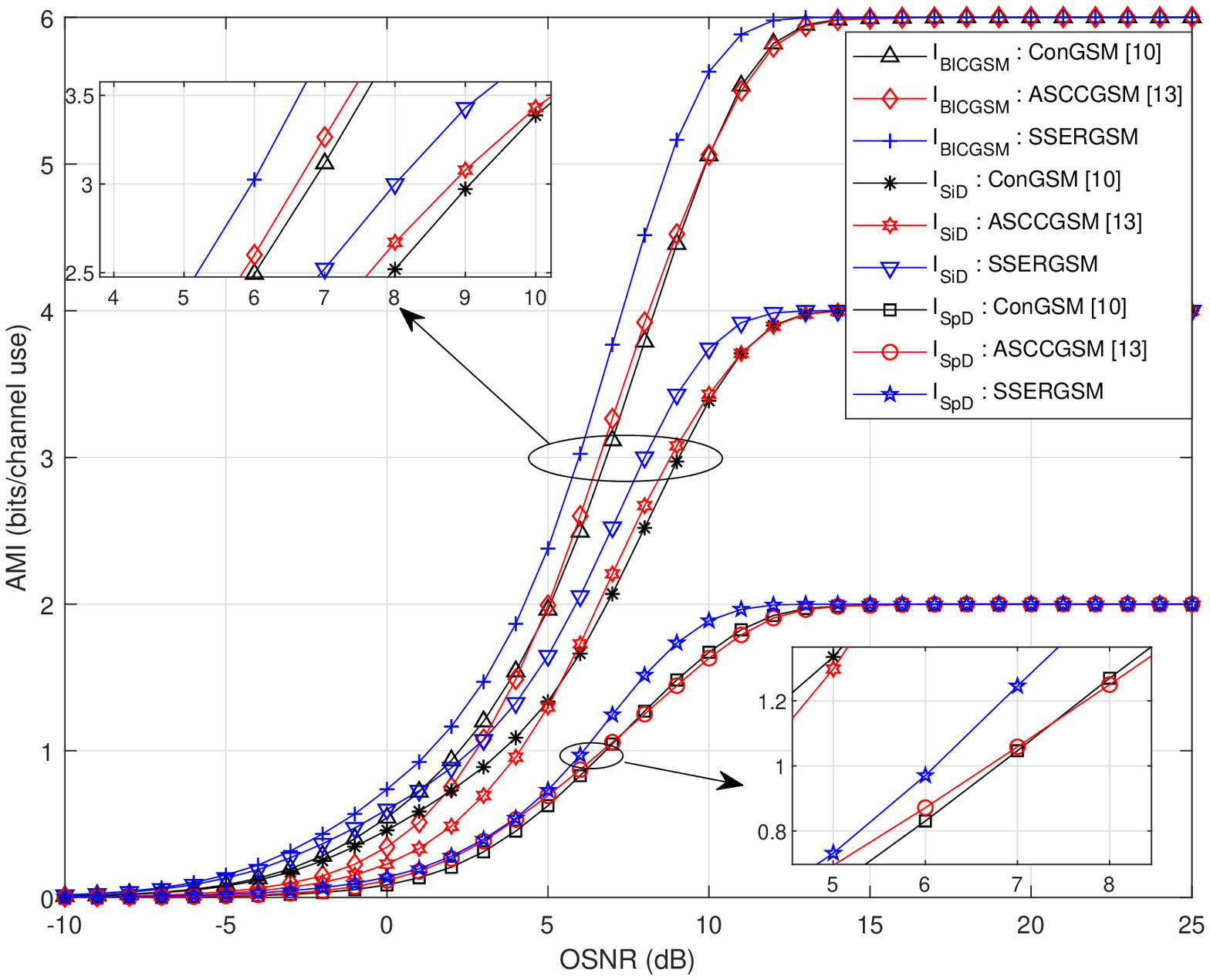}}\vspace{-2mm}
\caption{$I_{\rm BICGSM}$, $I_{{\rm SpD}}$ and $I_{{\rm SiD}}$ for the proposed SSERGSM, ConGSM, and ASCCGSM over a MIMO-VLC channel with $d_{\rm tx} = 0.5{\rm m}$, where the numbers of coded bits at each transmission instant are set as: (a)  $\rho = 4$ (i.e., $M = 2$) and (b) $\rho = 6$ (i.e., $M = 4$).}
\label{fig:dtx5-AMI} 
\end{figure}

According to Sect.~\ref{sect:section-2-C}, the achievable BICGSM-AMI (i.e., $I_{{\rm BICGSM}}$), SpD-AMI (i.e., $I_{{\rm SpD}}$) and SiD-AMI (i.e., $I_{{\rm SiD}}$) for a given GSM constellation can be measured. Assuming that the numbers of coded bit at each transmitted instant are set as $\rho = 4$ (i.e., $M = 2$) and $\rho = 6$ (i.e., $M = 4$), the AMIs for the proposed SSERGSM, ConGSM \cite{7416970}, and ASCCGSM \cite{7902182}
over a MIMO-VLC channel with $d_{\rm tx} = 0.5{\rm m}$ are depicted in Fig.~\ref{fig:dtx5-AMIa} and Fig.~\ref{fig:dtx5-AMIb}, respectively.
As observed from Fig.~\ref{fig:dtx5-AMIa}, the proposed SSERGSM exhibits
a significantly larger $I_{{\rm BICGSM}}$ with respect to the ConGSM and ASCCGSM. Meanwhile, $I_{{\rm SpD}}$ and $I_{{\rm SiD}}$ of the proposed SSERGSM are larger than their corresponding counterparts. Therefore, it can be concluded that the BICGSM scheme with the proposed SSERGSM mapper can obtain better performance compared with ConGSM and ASCCGSM mappers. Similar phenomenon also occurs in the case of $\rho = 6$ (i.e., $M = 4$).

\begin{figure}[tbp]
\center
\includegraphics[width=2.8in,height=2.05in]{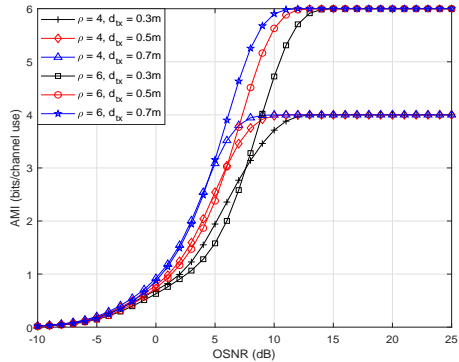}
\vspace{-3mm}
\caption{$I_{{\rm BICGSM}}$ for the proposed SSERGSM over MIMO-VLC channels with $d_{\rm tx} = 0.3{\rm m}$, $d_{\rm tx} = 0.5{\rm m}$ and $d_{\rm tx} = 0.7{\rm m}$, where the cases of $\rho = 4$ (i.e., $M = 2$) and $\rho = 6$ (i.e., $M = 4$) are considered.
\label{fig:Vdtx-AMI}}
\vspace{-3mm} 
\end{figure}

In order to show the influence of the spacing between two neighboring LEDs on the BICGSM capacity, we set three different values of $d_{\rm tx}$, i.e., $d_{\rm tx} = 0.3{\rm m}, 0.5{\rm m}, 0.7{\rm m}$ in the cases of $\rho = 4$ (i.e., $M = 2$) and $\rho = 6$ (i.e., $M = 4$). Then, the corresponding BICGSM capacities for the proposed SSERGSM are illustrated in Fig.~\ref{fig:Vdtx-AMI}. One can observe that
the proposed SSERGSM has the largest BICGSM capacity when $d_{\rm tx} = 0.7{\rm m}$, while it can achieve the lowest BICGSM capacity when $d_{\rm tx} = 0.3{\rm m}$ whether the case of $\rho = 4$ or $\rho = 6$. This indicates that the proposed SSERGSM with $d_{\rm tx} = 0.7{\rm m}$ can achieve the best performance among the three different cases of $d_{\rm tx}$.

\subsection{Convergence-Performance Analysis}\label{sect:section-3-C}
\subsubsection{Decoding Thresholds}\label{sect:section-3-C-1}
In order to further demonstrate the merit of the proposed SSERGSM in the BICGSM-ID MIMO-VLC systems, we conduct the convergence-performance analysis by means of the MPEXIT algorithm.
Suppose that all information bits
used in the above scenarios are encoded by the rate-$1/2$ AR$4$JA code \cite{5174517}. Unless otherwise stated, the information block length is $3600$ bits, and the maximum numbers of inner and outer iterations are assumed as $G_{1} = 20$ and $G_{2} = 4$, respectively. Furthermore, we consider three different values of $d_{\rm tx}$, i.e., $d_{\rm tx} = 0.3{\rm m}, 0.5{\rm m}, 0.7{\rm m}$, in the BICGSM-ID MIMO-VLC systems, where the cases of $\rho = 4$ and $\rho = 6$ are involved.

In Table~\ref{tab:4}, we analyze the decoding thresholds of rate-$1/2$ AR$4$JA code \cite{5174517} in the BICGSM-ID MIMO-VLC systems with
three different GSM constellations, i.e., the proposed SSERGSM, ConGSM \cite{7416970} and ASCCGSM \cite{7902182}.
As can be seen from Table~\ref{tab:4}, in the case of $\rho = 4$, the P-LDPC-coded BICGSM-ID MIMO-VLC system with the proposed SSERGSM constellation achieves the smaller decoding threshold compared with other two GSM constellations, i.e.,  ConGSM \cite{7416970} and ASCCGSM \cite{7902182} constellations, when $d_{\rm tx}$ are respectively set to be $0.3{\rm m}$, $0.5{\rm m}$ and $0.7{\rm m}$.
Moreover, the decoding thresholds of the AR$4$JA-coded BICGSM-ID schemes with the three different GSM constellations are reduced as $d_{\rm tx}$ varies from $0.3$m to $0.7$m, which indicates that system performance is optimal when $d_{\rm tx} = 0.7{\rm m}$. Similar conclusions can be drawn from the case of $\rho = 6$. We have also analyzed the decoding thresholds for the proposed SSERGSM constellation in the AR$4$JA-coded BICMGSM systems without ID, and have found that the ID framework converges faster than the non-ID framework.
Thus, the SSERGSM constellation can obtain better performance than other two GSM constellations in the P-LDPC-coded BICGSM-ID MIMO-VLC systems.

\begin{table}[t]\scriptsize
\caption{Decoding thresholds (dB) of the AR$4$JA-coded BICGSM-ID schemes with three different GSM constellations over MIMO-VLC channels with three different values of $d_{\rm tx}$, where the cases of $\rho = 4$ and $\rho = 6$ are considered. The maximum numbers of inner and outer iterations are assumed as $G_{1} = 20$ and $G_{2} = 4$, respectively.}
\centering
\begin{tabular}{|c|c|c|c|c|c|c|}
\hline
\backslashbox{$d_{\rm tx}$ (m)}{$\rho=4$} & SSERGSM  &  ConGSM \cite{7416970}  &   ASCCGSM \cite{7902182}\\
\hline
$0.3$  &  $5.731$  & $9.225$   &    $8.623$\\
\hline
$0.5$  &  $4.746$  &  $6.846$   &   $6.472$\\
\hline
$0.7$  &  $3.804$  &  $5.322$   &   $5.215$\\
\hline
\hline
\backslashbox{$d_{\rm tx}$ (m)}{$\rho=6$} & SSERGSM  &  ConGSM \cite{7416970} &  ASCCGSM \cite{7902182} \\
\hline
$0.3$  &  $7.603$  &  $10.612$   &   $10.027$\\
\hline
$0.5$  &  $6.585$  &  $8.324$   &    $7.749$\\
\hline
$0.7$  &  $5.278$  & $6.196$   &   $6.014$\\
\hline
\end{tabular}
\label{tab:4}
\end{table}

\subsubsection{Extrinsic MI of GSM Demapper}\label{sect:section-3-C-2}
\begin{figure}[tbp]
\centering
\subfigure[\hspace{-0.5cm}]{
\includegraphics[width=2.8in,height=2.15in]{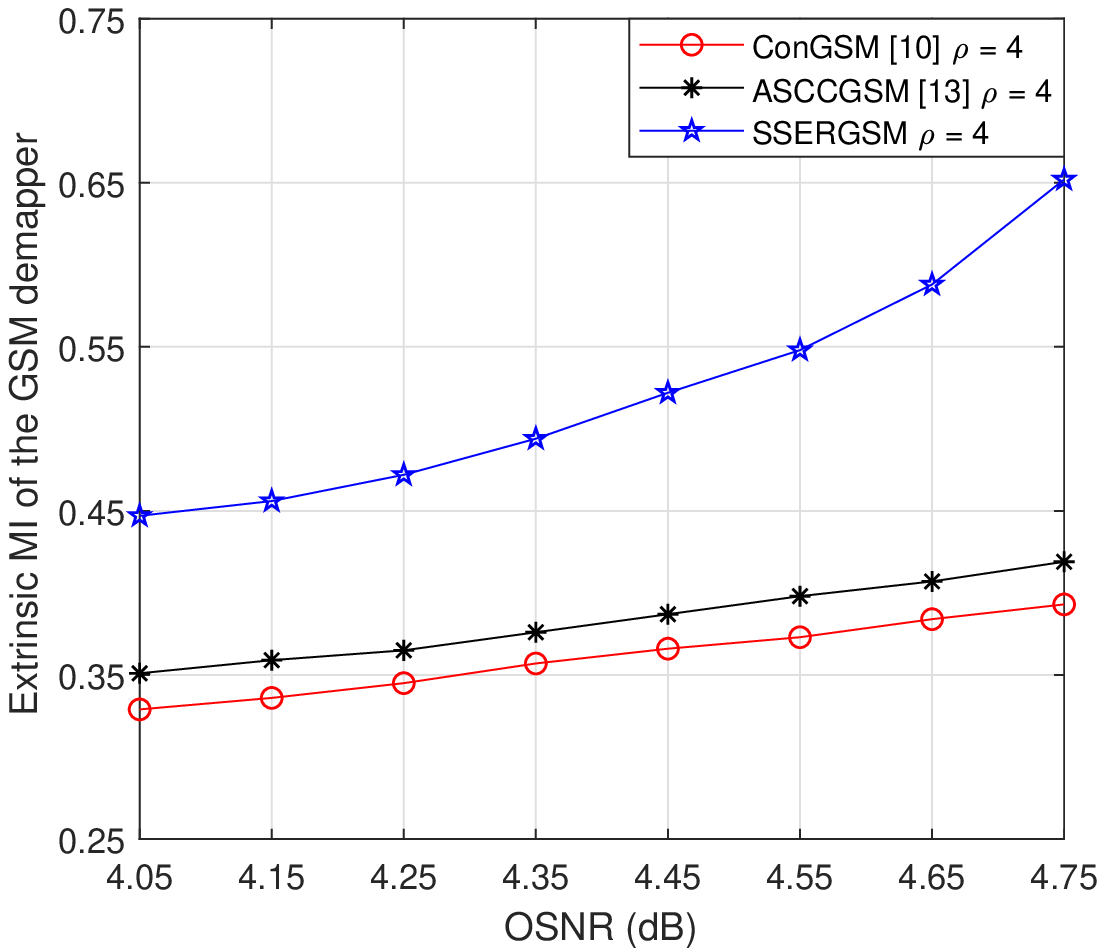}}
\subfigure[\hspace{-0.5cm}]{
\includegraphics[width=2.8in,height=2.15in]{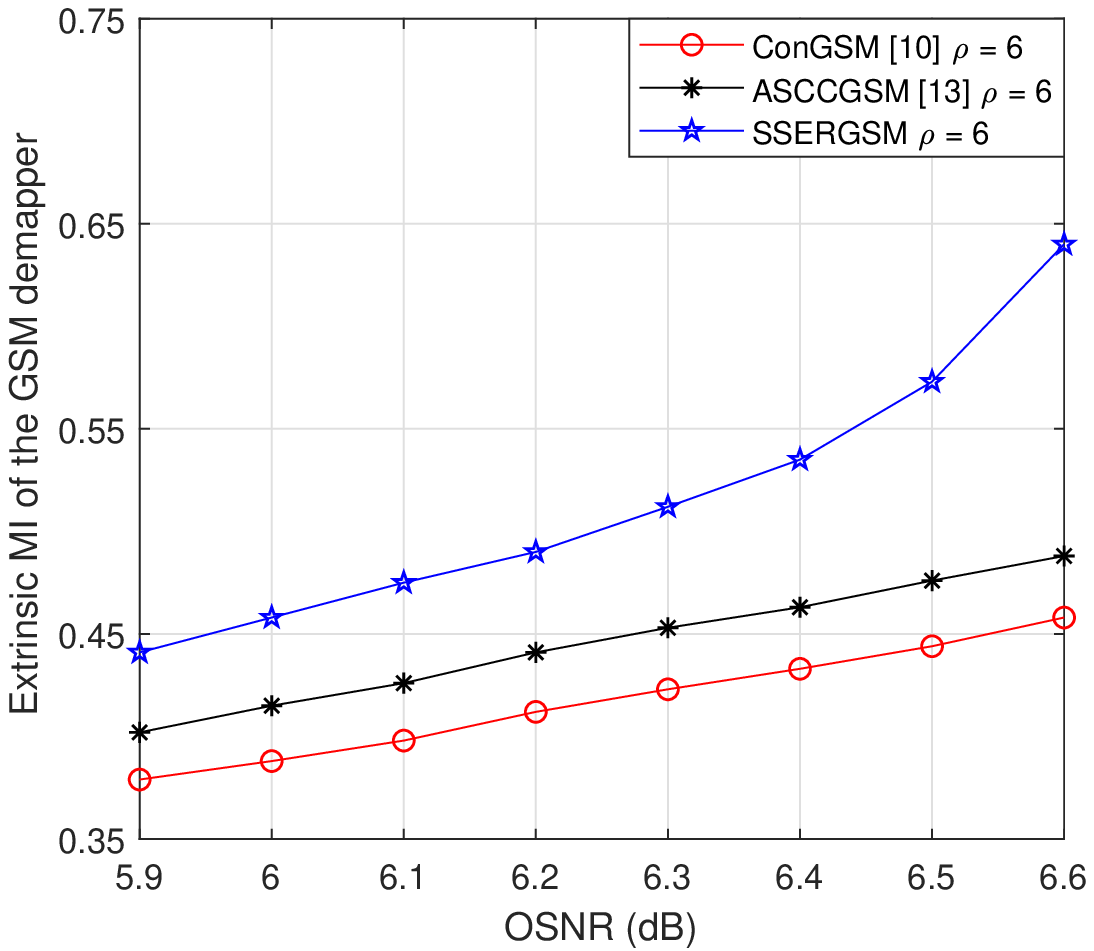}}\vspace{-1.5mm}
\caption{Extrinsic MIs for the proposed SSERGSM, ConGSM and ASCCGSM in a P-LDPC-coded BICGSM-ID MIMO-VLC system with $d_{\rm tx} = 0.5{\rm m}$, where the numbers of coded bits at each transmission instant are set as: (a)  $\rho = 4$ (i.e., $M = 2$) and (b) $\rho = 6$ (i.e., $M = 4$).}
\label{fig:dtx5-eMI}  \vspace{-5mm}
\end{figure}

To elaborate a little further, we analyze the extrinsic MIs output from the demappers corresponding to three different GSM constellations in the cases of $\rho = 4$ and $\rho = 6$ for the AR4JA-coded BICGSM-ID MIMO-VLC systems with $d_{\rm tx} = 0.5{\rm m}$, and show the results in Fig.~\ref{fig:dtx5-eMI}. As observed, in the case of $\rho = 4$ (see Fig.~\ref{fig:dtx5-eMI}(a)), when OSNR is varying from $4.05$ dB to $4.75$ dB, the proposed SSERGSM constellation obtains greater extrinsic MI compared with other two GSM constellations, implying that using the SSERGSM constellation can provide more reliable information for the demapper during the iterative process with respect to other two GSM constellations. It is well known that more reliable demapping information can accelerate the convergence of the P-LDPC-coded BICM-ID scheme. Similar conclusion can be found in Fig.~\ref{fig:dtx5-eMI}(b) with the case of $\rho = 6$. It can be easily observed that the extrinsic-MI analyses agree well with their corresponding decoding-threshold analyses. Therefore, given an OSNR range, the extrinsic MI of the GSM demapper can serve as a useful performance metric of a GSM constellation over the MIMO-VLC channels.

\section{Design of P-LDPC Codes for BICGSM-ID MIMO-VLC systems}
In order to improve the error correction performance of the BICGSM-ID MIMO-VLC systems, a type of rate-compatible P-LDPC codes is designed in this section.
\subsection{Design of P-LDPC Codes}
It is well known that the conventional AR4JA code \cite{5174517} has excellent error performance over the AWGN channels because it enjoys two desirable properties, i.e., low decoding threshold and linear minimum distance growth (the minimum Hamming distance growing linearly with the codeword
length) \cite{6266764}.  The base matrix corresponding to a rate-$(e+1)/(e+2)$ AR4JA code is given by
\begin{equation}
\bB_{\rm{AR4JA}} =
\begin{bmatrix}
\begin{array}{cc}
\begin{array}{ccccc}
1 & 2 & 0 & 0 & 0    \\
0 & 3 & 1 & 1 & 1    \\
0 & 1 & 2 & 2 & 1    \\
\end{array}
\overbrace{\begin{array}{ccccc}
0 & 0 & \cdots & 0 & 0 \\
1 & 3 & \cdots & 1 & 3 \\
3 & 1 & \cdots & 3 & 1 \\
\end{array}}^{2e}
\end{array}
\end{bmatrix},
\label{eq:14}
\end{equation}
where $e$ is the number of extension patterns (i.e., $e = 0,1,2,\cdots$), the matrix size is $3\times(2e+5)$, and the degree-$6$ variable node (i.e., the second column of $\mathbf{B}$) is punctured.

Typically, the performance of P-LDPC codes depends on not
only the structure of codes, but also the type of channel. The conventional P-LDPC codes, such as AR$4$JA code, that performs well over AWGN channels are not suitable for the usage in the BICGSM-ID MIMO-VLC systems.
Besides, by using the PEXIT analysis, we found that the P-LDPC codes with the lowest-degree variable node punctured can achieve better convergence performance with respect to the counterparts with other puncturing rules in the BICGSM-ID MIMO-VLC systems. Therefore, we aim to design a type of rate-compatible P-LDPC code with puncturing the lowest degree variable node.

\begin{figure}[t]
\center
\includegraphics[width=2.0in,height=2.15in]{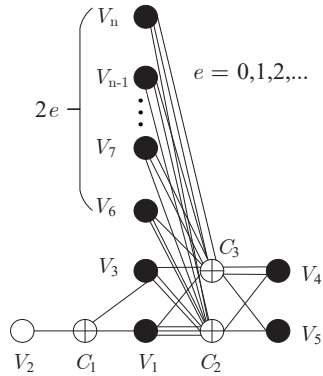}\vspace{-4mm}
\caption{Protograph structure of the proposed EARA P-LDPC code}
\label{fig:Fig.6}
\end{figure}
Considering a given codeword length, the lifting factor (i.e., ``copy-and-permuting'' times) for P-LDPC code will becomes smaller if the protograph
size increases, which makes it more difficult to avoid error floor. Furthermore, the larger the protograph is, the more complex the corresponding optimization algorithm will be. Therefore, we first consider a rate-$1/2$ mother base matrix with the same size as that of the conventional AR4JA code, which includes a punctured variable node. To facilitate the code design, we impose some constraints on the rate-$1/2$ P-LDPC code in order to guarantee that its mother base matrix possesses the linear-minimum-distance-growth property and a relatively low decoding threshold, as follows.

1) Initialize a protograph with a degree-$1$ variable node being the punctured variable node, which corresponds to the second column in the mother base matrix.

2) Impose empirical constraints for the linear-minimum-distance-growth property on the base matrix, i.e., only one degree-2 variable node is allowed in the mother base matrix.

3) Set the maximum number of parallel edge to be $3$ in the mother base matrix so as to maintain the low encoding and decoding complexity. This is because the encoding and decoding complexity of a P-LDPC code can be reflected by the number of edges in its corresponding protograph. Besides, for a fixed codeword length, if the number of parallel edges increases, the likelihood of short cycles will increase.

Based on the above rules, the mother base-matrix structure of the proposed rate-$1/2$ P-LDPC code can be formulated as

\begin{equation}
\mathbf{B}_{1/2}=
\begin{bmatrix}
   \begin{array}{lllllll}
     b_{1,1}  &   1  &   b_{1,3}   &   b_{1,4}     &  0      \\
    b_{2,1}  &   0  &   b_{2,3}   &   b_{2,4}     &  1      \\
    b_{3,1}  &  0  &   b_{3,3}   &   b_{3,4}     &  1    \\

   \end{array}
\end{bmatrix}.
\end{equation}

To assign the first column in the mother base matrix $\mathbf{B}_{1/2}$ as the largest degree variable node, we add an additional constraint, i.e., $b_{1,1} + b_{2,1} + b_{3,1} > b_{1,3} + b_{2,3} + b_{3,3}$ and $b_{1,1} + b_{2,1} + b_{3,1} > b_{1,4} + b_{2,4} + b_{3,4}$.
After a exhaustive search using the proposed MPEXIT algorithm, one can obtain the optimal mother base matrix with lowest decoding threshold and linear-minimum-distance-growth property, as

\begin{equation}
\mathbf{B}_{1/2}=
\begin{bmatrix}
   \begin{array}{lllllll}
    1  &   1  &   1   &   0     &  0      \\
    3  &   0  &   2   &   1     &  1      \\
    1  &   0  &   1   &   2     &  1    \\
   \end{array}
\end{bmatrix}.
\end{equation}

However, for higher-rate P-LDPC code design, the search space becomes larger, hence the corresponding exhaustive-search method is more complex.
Therefore, we resort to a pattern-extension method, which is similar to the construction of the AR4JA codes, to design the higher-rate P-LDPC code. In particular, the pattern-extension method can be realized by repeatedly
adding two variable nodes into the resultant mother protograph (i.e., base matrix).
It has been demonstrated in \cite{5174517,7112076} that adding degree-3 variable node into a rate-
$1/2$ protograph can maintain the lowest-complexity feature for
its corresponding higher-rate counterparts without deteriorating
the linear-minimum-distance-growth property.

For the above reason, a type of enhanced rate-compatible P-LDPC codes with the code rate $R = (e+1)/(e+2)$ ($e = 0,1,2,\cdots$), referred to {\em EARA codes}, is obtained. The protograph structure of the proposed EARA code
is illustrated in Fig.~\ref{fig:Fig.6}, where the dark circles denote transmitted variable nodes, the white circle denotes the punctured variable node, and the plus circle are check nodes. As a result, the corresponding base matrix is given as
\begin{equation}
\bB_{\rm{EARA}} =
\begin{bmatrix}
\begin{array}{cc}
\begin{array}{ccccc}
1 & 1 & 1 & 0 & 0    \\
3 & 0 & 2 & 1 & 1    \\
1 & 0 & 1 & 2 & 1    \\
\end{array}
\overbrace{\begin{array}{ccccc}
0 & 0 & \cdots & 0 & 0 \\
2 & 1 & \cdots & 2 & 1 \\
1 & 2 & \cdots & 1 & 2 \\
\end{array}}^{2e}
\end{array}
\end{bmatrix}.
\end{equation}

\subsection{Asymptotic Performance Metrics}
\subsubsection{Decoding-Threshold Analysis}
To verify the convergence performance of the proposed EARA P-LDPC code, we take the rate-$2/3$ AR$4$JA code \cite{5174517}, accumulate-repeat-by-$4$-accumulate (AR$4$A) code \cite{7320948}, improved-AR$4$JA (IAR$4$JA) code, improved-AR$4$A (IAR$4$A) code and regular-$(3,9)$ code \cite{5174517} as benchmarks in the proposed SSERGSM-aided BICGSM-ID MIMO-VLC systems with $d_{\rm tx} = 0.3{\rm m}$, and show their corresponding decoding thresholds in Table~\ref{tab:tab6}.
It is noted that the IAR4JA code and IAR4A code respectively represent two modified P-LDPC codes possessing the same base matrix as the AR$4$JA code and the AR$4$A code, but only with puncturing the lowest degree variable nodes. As shown, the proposed EARA P-LDPC code possesses the lowest decoding
threshold among the six types of P-LDPC codes in the cases of $\rho = 4$ and $\rho = 6$. This implies that the proposed EARA P-LDPC code can achieve better error performance in low-OSNR region compared with its counterparts in the proposed SSERGSM-aided BICGSM-ID MIMO-VLC systems with $d_{\rm tx} = 0.3{\rm m}$.

\begin{table*}[t]\scriptsize
\center
\caption{Decoding thresholds of P-LDPC codes, i.e., the proposed EARA code, AR4JA code, AR4A code, IAR4JA code, IAR4A code and regular-$(3,9)$ code, in the proposed SSERGSM-aided BICGSM-ID MIMO-VLC systems with $d_{\rm tx} = 0.3{\rm m}$, where $\rho = 4$ and $\rho = 6$ are considered. The maximum numbers of inner and outer iterations are assumed as $G_{1} = 20$ and $G_{2} = 4$, respectively.}
\begin{tabular}{|c|c|c|c|c|c|c|c|}
\hline
$\rho = 4$        &  EARA        &  AR$4$JA      &AR$4$A &  IAR$4$JA     & IAR$4$A  &  Regular \\
\hline
${\rm OSNR}$(dB)      &  $5.314$    &  $7.226$     & $7.145$  &  $6.214$  & $6.157$  & $6.012$  \\
\hline\hline
$\rho = 6$       &  EARA     &  AR$4$JA     &AR$4$A  &  IAR$4$JA    & IAR$4$A   &  Regular \\
\hline
${\rm OSNR}$(dB)     &  $7.108$     &  $8.625$   &$8.512$  &  $8.028$   & $7.964$    &  $7.813$ \\
\hline
\end{tabular}
\label{tab:tab6}
\end{table*}


\subsubsection{AWD Analysis}
By exploiting AWD function \cite{5174517,7112076}, we measure the TMDRs for rate-$2/3$ P-LDPC codes, i.e., the proposed EARA code, AR$4$JA code, AR$4$A code, IAR$4$JA code, IAR$4$A code and regular-$(3,9)$ code and give the corresponding results in Table~\ref{tab:tab7}. Referring to this table, except that the AR$4$A code and IAR$4$A code, the other four types of P-LDPC codes possess effective TMDR values (i.e., positive TMDR values).

Based on the above decoding-threshold and AWD analyses, it is affirmed that the proposed EARA code not only possesses the lowest threshold but also exhibits a linear-minimum-distance property.
This demonstrates that the proposed EARA code achieves better convergence performance compared with its counterparts in the low-OSNR region and provides a desirable error performance in the high-OSNR region in the proposed SSERGSM-aided BICGSM-ID MIMO-VLC systems.

{\em Remark $2$:} We have also evaluated the six types of P-LDPC codes with other code rates (e.g., $R = 1/2, 3/4$) and have observed that the EARA codes also obtain the lowest decoding threshold and have effective TMDR values.

\begin{table}[tp]
\center
\caption{TMDRs of rate-$2/3$ P-LDPC codes, i.e., the proposed EARA code, AR$4$JA code, AR$4$A code, IAR$4$JA code, IAR$4$A code and the regular-$(3,9)$ code, in the BICGSM-ID MIMO-VLC systems.}
\begin{tabular}{|c|c|c|c|c|c|c|}
\hline
Code Type       &  EARA        &  AR$4$JA      &AR$4$A &  IAR$4$JA     & IAR$4$A  &  Regular \\
\hline
TMDR      &  $0.004$    &  $0.009$    & N.A   &  $0.008$  & N.A  & $0.006$ \\
\hline
\end{tabular}
\label{tab:tab7}
\end{table}

\subsection{Complexity Analysis}
In this subsection, we analyze the computational complexity of the P-LDPC codes for successful decoding under the serially concatenated ID framework. In this ID framework, GSM demapper and P-LDPC decoder impose the main computational complexity. Therefore, based on the log-domain maximum a-posteriori
probability (Log-Map) demapping algorithm \cite{123456} and the LLR-based belief
propagation (LLR-BP) decoding algorithm \cite{1495850}, we evaluate the numbers of real addition (RA) and real multiplication (RM) operations of a P-LDPC code for successful decoding under the ID framework. The results are shown in Table~\ref{tab:complexity}, where $n$ and $m$ are the number of VNs and CNs, respectively; $p$ is the number of punctured VNs; $\bar{g}_{\rm v}$ and $\bar{g}_{\rm c}$ denote the average degrees of VNs and CNs, respectively; $\bar{T}_{\rm 1}$ and $\bar{T}_{\rm 2}$ denote the average numbers of inner and outer iterations. As a further investigation, Table~\ref{tab:specially-complexity} presents the average numbers of RA and RM operations of different P-LDPC codes for successful decoding in the SSERGSM-aided BIGSM-ID MIMO-VLC systems with $d_{\rm tx} = 0.3{\rm m}$ and $\rho=4$. As can be seen from this table, the proposed EARA P-LDPC code possesses a lower computational complexity compared with its counterparts under the ID framework.

\begin{table*}[tbp]
\center
\caption{Average number of operations of a P-LDPC code for successful decoding under the ID framework.}
\begin{tabular}{|c|c|c|}
\hline
\backslashbox{Component}{Operation} & RA  &  RM  \\
\hline
Log-Map demapping  &  $(n-p)[2^{\rho}((N_{\rm t}+1)N_{\rm r} + 2\rho -4)+2]{\bar{T}_{\rm 2}}$  & $2^{\rho}(n-p)[(N_{\rm t}+1)N_{\rm r} + \rho +2]{\bar{T}_{\rm 2}}$   \\
\hline
LLR-BP decoding  &  $m(\bar{g}_{\rm v}-1)\bar{T}_{\rm 1} $ &  $2n\bar{g}_{\rm c}\bar{T}_{\rm 1}$ \\
\hline
\end{tabular}\vspace{-2mm}
\label{tab:complexity}
\end{table*}

\section{Simulation Results}\label{sect:section-6}
To validate our proposed SSERGSM constellation and constructed EARA P-LDPC code, we present various simulation results on the BICGSM-ID MIMO-VLC systems in this section. To be specific, a $4\times4$ indoor LOS MIMO-VLC channel model is considered. Furthermore, we assume that the information block length of P-LDPC codes is $3600$ bits. In addition, unless otherwise stated, the maximum numbers of inner iterations and outer iterations are $G_{1} = 20$ and $G_{2} = 4$, respectively.

\subsection{BER Performance of Proposed SSERGSM Constellation }\label{sect:section-6-A}
\begin{figure}[t]
\centering
\subfigure[\hspace{-0.5cm}]{ \label{fig:dtx5-BERa}
\includegraphics[width=3.5in,height=2.5in]{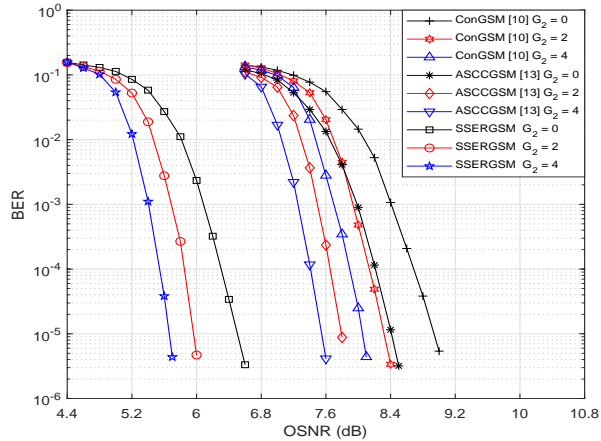}}
\subfigure[\hspace{-0.5cm}]{ \label{fig:dtx5-BERb}
\includegraphics[width=3.5in,height=2.5in]{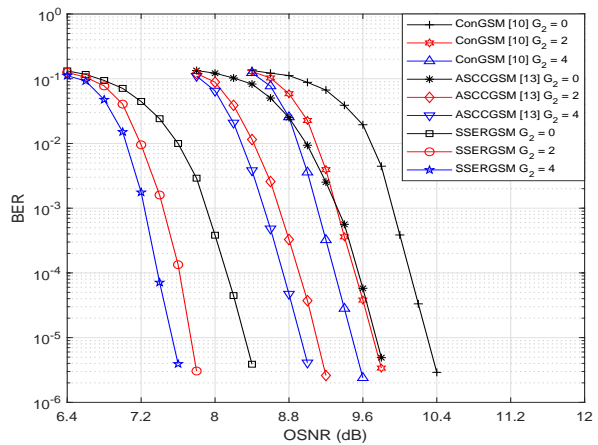}}
\caption{BER curves of the AR$4$JA-coded BICGSM-ID schemes with the proposed SSERGSM, ASCCGSM, and ConGSM constellations over a MIMO-VLC channel with $d_{\rm tx} = 0.5{\rm m}$, where (a) $\rho = 4$ (i.e., $M = 2$) and (b) $\rho = 6$ (i.e., $M = 4$) are considered. The maximum numbers of outer iterations are set as $G_{2} = 0,2,4$.}
\label{fig:dtx5-BER}
\vspace{-2mm}
\end{figure}

Fig.~\ref{fig:dtx5-BER} depicts the bit-error-rate (BER) curves of rate-$1/2$ the AR$4$JA-coded BICGSM-ID schemes with three different types of GSM constellations, i.e., the proposed SSERGSM constellations, ConGSM \cite{7416970}, and ASCCGSM \cite{7902182} over a MIMO-VLC channel with $d_{\rm tx} = 0.5{\rm m}$, where the maximum numbers of outer iterations are set as $G_{2} = 0,2,4$.
For the case of $\rho = 4$ (i.e., $M = 2$), as observed from Fig.~\ref{fig:dtx5-BERa},
the proposed SSERGSM constellation can achieve better error performance compared with the two counterparts in the BICGSM-ID MIMO-VLC systems. To be specific, when $G_{2} = 4$, the proposed SSERGSM constellation exhibits remarkable gains of about $1.9$ dB and $2.4$ dB over the ASCCGSM and the ConGSM constellations, respectively, at a BER of $7\times10^{-6}$. Furthermore, when $G_{2} = 0$ (i.e., BICGSM system), it is evident that the proposed SSERGSM constellation can also obtain larger gains compared with its counterparts,
indicating that the proposed SSERGSM constellation is suitable for not only ID scenario but also non-ID scenario.
Additionally, it can be easily observed that, at a BER $5\times10^{-6}$, the proposed SSERGSM constellation with $G_{2} = 4$ obtains about $0.3$-dB gain over that with $G_{2} = 2$, while the latter obtains about $0.56$-dB gain over that with $G_{2} = 0$. 
Similar observations can be also found in the case of $\rho = 6$ (i.e., $M = 4$) (see Fig.~\ref{fig:dtx5-BERb}).
Based on the above discussions, the proposed SSERGSM constellation will become better as the maximum number of out iterations (i.e., the value of $G_2$) increases, while the relative performance between the proposed SSERGSM constellation and the benchmarks remains the same.

\begin{figure}[t]
\centering
\subfigure[\hspace{-0.5cm}]{
\includegraphics[width=3.5in,height=2.5in]{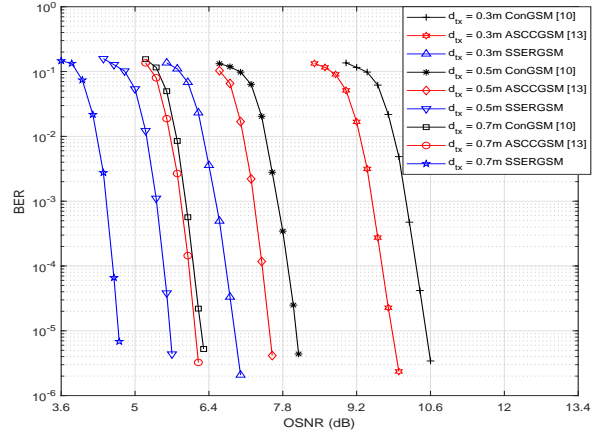}}
\subfigure[\hspace{-0.5cm}]{
\includegraphics[width=3.5in,height=2.5in]{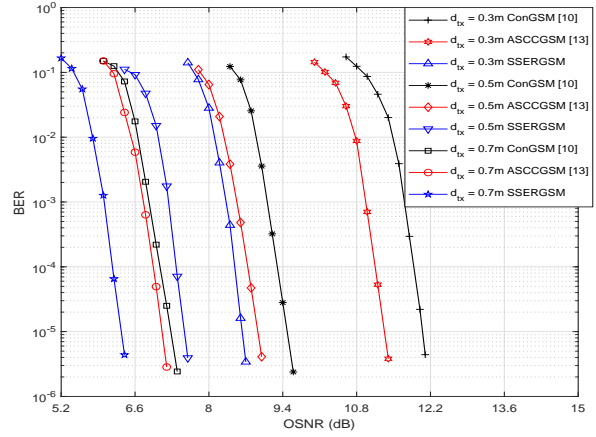}}
\caption{BER curves of the AR$4$JA-coded BICGSM-ID schemes with the proposed SSERGSM, ASCCGSM, and ConGSM constellations over MIMO-VLC channels with (a) $\rho = 4$ (i.e., $M = 2$) and (b) $\rho = 6$ (i.e., $M = 4$). The spacings between two neighboring LEDs of are set as $d_{\rm tx} = 0.3{\rm m},0.5{\rm m},~{\rm and}~0.7{\rm m}$.}
\vspace{-2mm}
\label{fig:Fig.8}
\end{figure}

As a further investigation, we perform the BER performance for three different types of GSM constellations, i.e., the proposed SSERGSM constellations, ConGSM \cite{7416970}, ASCCGSM \cite{7902182} in the AR4JA-coded BICGSM-ID MIMO-VLC systems with three different values of $d_{\rm tx}$ (i.e., $d_{\rm tx} = 0.3{\rm m},0.5{\rm m},~{\rm and}~0.7{\rm m}$) in
Fig.~\ref{fig:Fig.8}. Referring to Fig.~\ref{fig:Fig.8}(a) corresponding to the case of $\rho = 4$ (i.e., $M = 2$), at a
BER of $7\times10^{-6}$, the proposed SSERGSM constellation only requires about a $4.8$-dB OSNR when $d_{\rm tx} = 0.7{\rm m}$, while it needs about $5.68$-dB and $6.88$-dB OSNR when $d_{\rm tx} = 0.5{\rm m}$ and $0.3{\rm m}$, respectively.
We also observe that, aiming to achieve a BER of $7\times10^{-6}$, the proposed SSERGSM constellation benefits from performance gains of about $3$ dB and $3.64$ dB compared with the ASCCGSM and ConGSM constellations when $d_{\rm tx} = 0.3{\rm m}$. When $d_{\rm tx} = 0.5{\rm m}$, the proposed SSERGSM constellation also significantly outperforms the ASCCGSM and ConGSM constellations about $1.9$-dB and $2.4$-dB gains, respectively. When $d_{\rm tx} = 0.7{\rm m}$, this performance gains of the proposed SSERGSM constellation with respect to the ASCCGSM and ConGSM constellations are only about $0.45$ dB and $0.6$ dB, respectively. Likewise, one can observe from Fig.~\ref{fig:Fig.8}(b) that the relative performance among the three different types of GSM constellations in the case of $\rho = 6$ (i.e., $M = 4$) is identical to that in the case of $\rho = 4$ (i.e., $M = 2$).
\begin{table}[tbp]
\center
\caption{Average Numbers of RA and RM operations of different P-LDPC codes for successful decoding under the ID framework.}
\begin{tabular}{|c|c|c|}
\hline
\backslashbox{Code Type}{Operation}    & RA  &  RM  \\
\hline
EARA  &  $2.642\times10^{6}$  & $6.602\times10^{6}$   \\
\hline
AR$4$JA & $8.853\times10^{6}$  & $2.205\times10^{7}$  \\
\hline
AR$4$A & $8.762\times10^{6}$  & $2.173\times10^{7}$  \\
\hline
IAR$4$JA & $4.565\times10^{6}$  & $1.123\times10^{7}$  \\
\hline
IAR$4$A  & $4.241\times10^{6}$  & $1.025\times10^{7}$  \\
\hline
Regular & $2.819\times10^{6}$  & $7.789\times10^{6}$ \\
\hline
\end{tabular}
\label{tab:specially-complexity}
\end{table}
These phenomenons suggest that the P-LDPC-coded BICGSM-ID MIMO-VLC system performance is continuously improved as $d_{\rm tx}$ increases from $0.3{\rm m}$ to $0.7{\rm m}$, while the relative performance-gain between the proposed SSERGSM constellation and the benchmarks becomes smaller simultaneously.
The reason is the fact that the parameter $d_{\rm tx}$ plays an important role in affecting the correlation of the MIMO-VLC channel.
Specifically, a smaller $d_{\rm tx}$ produces a higher correlated MIMO-VLC channel.
It manifests that the proposed SSERGSM constellation has greater advantage in the highly-correlated MIMO-VLC channels.
\subsection{BER/FER Performance of Proposed EARA P-LDPC Code}\label{sect:section-6-B}
\begin{figure}[tbp]
\centering
\subfigure[\hspace{-0.5cm}]{
\includegraphics[width=3.5in,height=2.5in]{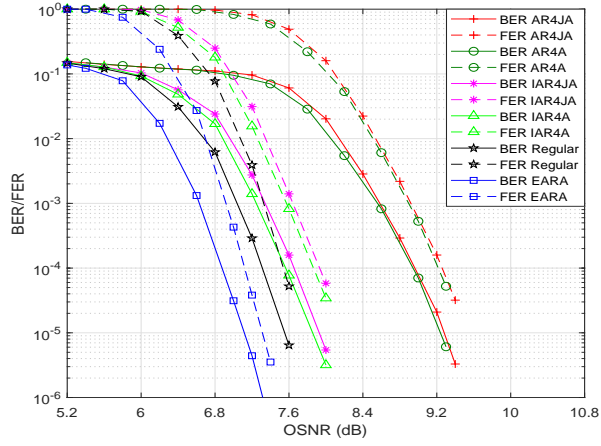}}
\subfigure[\hspace{-0.5cm}]{
\includegraphics[width=3.5in,height=2.5in]{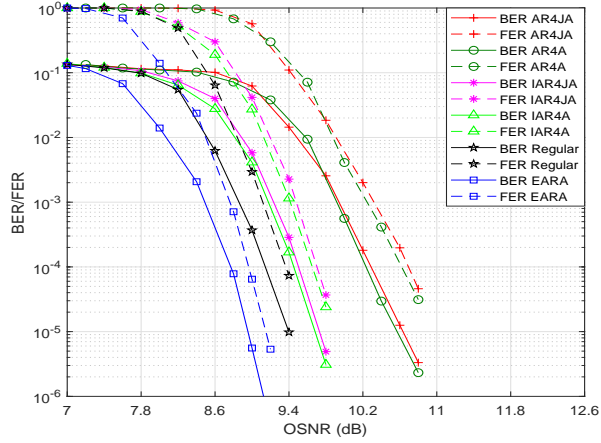}}
\caption{BER/FER curves of six types of P-LDPC codes in the SSERGSM-aided BICGSM-ID MIMO-VLC systems with $d_{\rm tx} = 0.3{\rm m}$, where (a) $\rho = 4$ (i.e., $M = 2$) and (b) $\rho = 6$ (i.e., $M = 4$) are considered.}
\vspace{-2mm}
\label{fig:Fig.9}
\end{figure}

Here, we carry out some BER/frame-error-rate (FER) performance simulations on the rate-$2/3$ proposed EARA code, the AR$4$JA code \cite{5174517}, AR$4$A code \cite{7320948}, IAR$4$JA code,
IAR$4$A code and the regular-$(3,9)$ code \cite{5174517} in the SSERGSM-aided BICGSM-ID MIMO-VLC systems with $d_{\rm tx} = 0.3 {\rm m}$, where the cases of $\rho = 4$ (i.e., $M = 2$) and $\rho = 6$ (i.e., $M = 6$) are considered. The results are illustrated in Fig.~\ref{fig:Fig.9}.
According to Fig.~\ref{fig:Fig.9}(a), at a BER of $7\times10^{-6}$, the proposed EARA P-LDPC code outperforms the IAR$4$A code and the IAR$4$JA code by about $0.67$ dB and $0.78$ dB, respectively. Moreover, the proposed EARA P-LDPC code achieves more than $0.4$-dB gain over the regular-$(3,9)$ code, while the regular-$(3,9)$ code further respectively performs better than the AR$4$A code and the AR$4$JA code by about $1.15$ dB and $1.69$ dB in the case of $\rho = 4$ (i.e., $M = 2$). Referring to Fig.~\ref{fig:Fig.9}(b) for the case of $\rho = 6$ (i.e., $M = 4$), at ${\textmd{OSNR}} = 9.0$ dB, the proposed EARA P-LDPC code can achieve a BER of $6\times10^{-6}$, while {\color{black}the AR$4$JA code, AR$4$A code, IAR$4$JA code, IAR$4$A code and the regular-$(3,9)$ code} accomplish BERs of $7.4\times10^{-2}$, $5.2\times10^{-2}$, $5.8\times10^{-3}$, $4.1\times10^{-3}$ and $3.7\times10^{-4}$, respectively. Importantly, there is no error-floor phenomenon for the proposed EARA P-LDPC code when the BER reduces to $10^{-6}$.
Based on the above observations, the proposed EARA P-LDPC code exhibits the best performance among the six types of P-LDPC codes.

{\em Remark $3$:} We have also carried out simulations with other system setups (e.g., different values of $N_{\rm t}$, $N_{\rm r}$, $N_{\rm a}$, $G_1$, $G_2$ and $\rho$), and have found that the relative performance between the proposed SSERGSM constellation and the benchmarks remains the same, which verifies the robustness of our proposed design.

\section{Conclusion}\label{sect:section-7}
In this paper, we presented a new BICGSM-ID transmission scheme over the MIMO-VLC channels. The design and optimization of GSM constellations and P-LDPC codes for use in the BICGSM-ID have been carefully investigated.
To be specific, we modified the PEXIT algorithm according to the characteristics of GSM and MIMO-VLC channel. Next, we put forward an SSER design method to formulate a type of SSERGSM constellations tailored for the P-LDPC-coded BICGSM-ID MIMO-VLC systems, which exhibits better performance than other GSM constellations in this scenario. By exploiting the MPEXIT algorithm, we also constructed a family of EARA P-LDPC codes, which possess lower decoding thresholds than other P-LDPC codes and desirable linear-minimum-distance-growth property in the SSERGSM-aided BICGSM-ID MIMO-VLC systems.
Simulation results validated that the joint deployment of SSERGSM constellations and EARA P-LDPC codes allow the BICGSM-ID MIMO-VLC systems to achieve very desirable error performance.
In addition, we analyzed the effect of the correlation of MIMO-VLC channel on the system performance.
Owing to these advantages, the proposed P-LDPC-coded BICGSM-ID is expected to promise a good solution for high-throughput MIMO-VLC applications.


\bibliographystyle{IEEEtran}  

\end{document}